\documentclass[journal]{vgtc}                

\ifpdf
  \pdfoutput=1\relax                   
  \usepackage{graphicx}                
  \DeclareGraphicsExtensions{.pdf,.png,.jpg,.jpeg} 
\else
  \usepackage{graphicx}                
  \DeclareGraphicsExtensions{.eps}     
\fi%

\graphicspath{{./figs/}{./figures/}{./pictures/}{./images/}{./}} 

\usepackage{amsmath}
\usepackage{microtype}                 
\PassOptionsToPackage{warn}{textcomp}  
\usepackage{textcomp}                  
\usepackage{mathptmx}                  
\usepackage{times}                     
\usepackage{cite}                      
\usepackage{tabu}                      
\usepackage{booktabs}                  

\usepackage{rotating}
\usepackage{makecell}
\usepackage{amsthm}
\usepackage{xspace}
\usepackage{subfigure}
\usepackage[usenames,dvipsnames,table]{xcolor}
\usepackage{boxedminipage}
\usepackage{algorithmic}
\usepackage[algoruled]{algorithm2e}
\usepackage{comment}
\usepackage{url}
\usepackage{color}
\usepackage{enumitem}
\usepackage{xfrac}
\usepackage{tabularx}
\usepackage{multirow}
\usepackage{booktabs}
\usepackage{amsmath,amssymb}
\usepackage{wrapfig}
\usepackage{flushend}

\newcommand{\hidecomment}[1]{}
\newcommand{\denselist}{\itemsep 0pt\parsep=1pt\partopsep 0pt}

\newcommand{\gustavo}[1]{{\color{red} Gustavo: [{#1}]}}
\newcommand{\harish}[1]{{\color{cyan} Harish: [{#1}]}}

\newcommand{\dataset}{data set\xspace}
\newcommand{\Dataset}{Data set\xspace}
\newcommand{\datasets}{data sets\xspace}

\newcommand{\squishlist}{
   \begin{list}{$\bullet$}
    { \setlength{\itemsep}{0pt}      \setlength{\parsep}{3pt}
      \setlength{\topsep}{3pt}       \setlength{\partopsep}{0pt}
      \setlength{\leftmargin}{1.5em} \setlength{\labelwidth}{1em}
      \setlength{\labelsep}{0.5em} } }

\newcommand{\squishlisttwo}{
   \begin{list}{$\bullet$}
    { \setlength{\itemsep}{0pt}    \setlength{\parsep}{0pt}
      \setlength{\topsep}{0pt}     \setlength{\partopsep}{0pt}
      \setlength{\leftmargin}{2em} \setlength{\labelwidth}{1.5em}
      \setlength{\labelsep}{0.5em} } }

\newcommand{\squishend}{
    \end{list}  }

\newcommand{\Rspace}        {{\mathbb R}}

\newcommand{\Kspace}        {{\mathbb K}}

\newtheorem{lemma}{Lemma}
\newtheorem{proposition}{Proposition}

\usepackage{eqparbox}

\newlength{\oldtextfloatsep}
\newlength{\oldcolumnsep}

\newcommand{\myparagraph}[1]{\noindent\textbf{#1}\xspace}

\onlineid{1049}

\vgtccategory{Research}
\vgtcpapertype{technique}

\title{TopoMap: A 0-dimensional Homology Preserving Projection of High-Dimensional Data}

\author{Harish Doraiswamy, Julien Tierny, Paulo J. S. Silva, Luis Gustavo Nonato,  and Claudio Silva}
\authorfooter{
\item H. Doraiswamy and C. Silva are with New York University;
J. Tierny is with CNRS and Sorbonne Universit\'e;
P. J. S. Silva is with University of Campinas; and
L. G. Nonato is with  University of Sao Paulo, Sao Carlos.
\item E-mail: \{harishd,csilva\}@nyu.edu, julien.tierny@sorbonne-universite.fr, pjssilva@ime.unicamp.br, gnonato@icmc.usp.br
}

\shortauthortitle{Doraiswamy \MakeLowercase{\textit{et al.}}: TopoMap: A 0-dimensional Homology Preserving Projection of High-Dimensional Data}

\abstract{
Multidimensional Projection is a fundamental tool for high-dimensional data analytics and visualization.
With very few exceptions, projection techniques are designed to map data from a high-dimensional space to a visual
space so as to preserve some dissimilarity (similarity) measure,
such as the Euclidean distance for example. In fact, although adopting distinct mathematical formulations designed to favor different aspects of the data,
most multidimensional projection methods strive to preserve dissimilarity 
measures that encapsulate geometric properties such as distances or
the proximity relation between data objects. However, geometric relations are 
not the only interesting property to be preserved in a projection.
For instance, the analysis of particular structures such as clusters and outliers could be more reliably performed if
the mapping process gives some guarantee as to topological invariants such as connected components and loops.
This paper introduces \emph{TopoMap}, 
a novel projection technique which provides
topological guarantees during the mapping process.
In particular, the
proposed method performs the mapping from a high-dimensional space to a 
visual space, while preserving the 0-dimensional persistence 
diagram of the Rips filtration of the high-dimensional data,
ensuring that the filtrations generate the same connected components when 
applied to the original as well as projected data.
The presented case studies show that the topological guarantee provided by TopoMap not only brings confidence to the visual analytic process but also
can be used to assist in the assessment of other projection methods.
} 


\CCScatlist{ 
 \CCScat{K.6.1}{Management of Computing and Information Systems}%
{Project and People Management}{Life Cycle};
 \CCScat{K.7.m}{The Computing Profession}{Miscellaneous}{Ethics}
}

\teaser{
 \centering
 \includegraphics[width=0.9\linewidth]{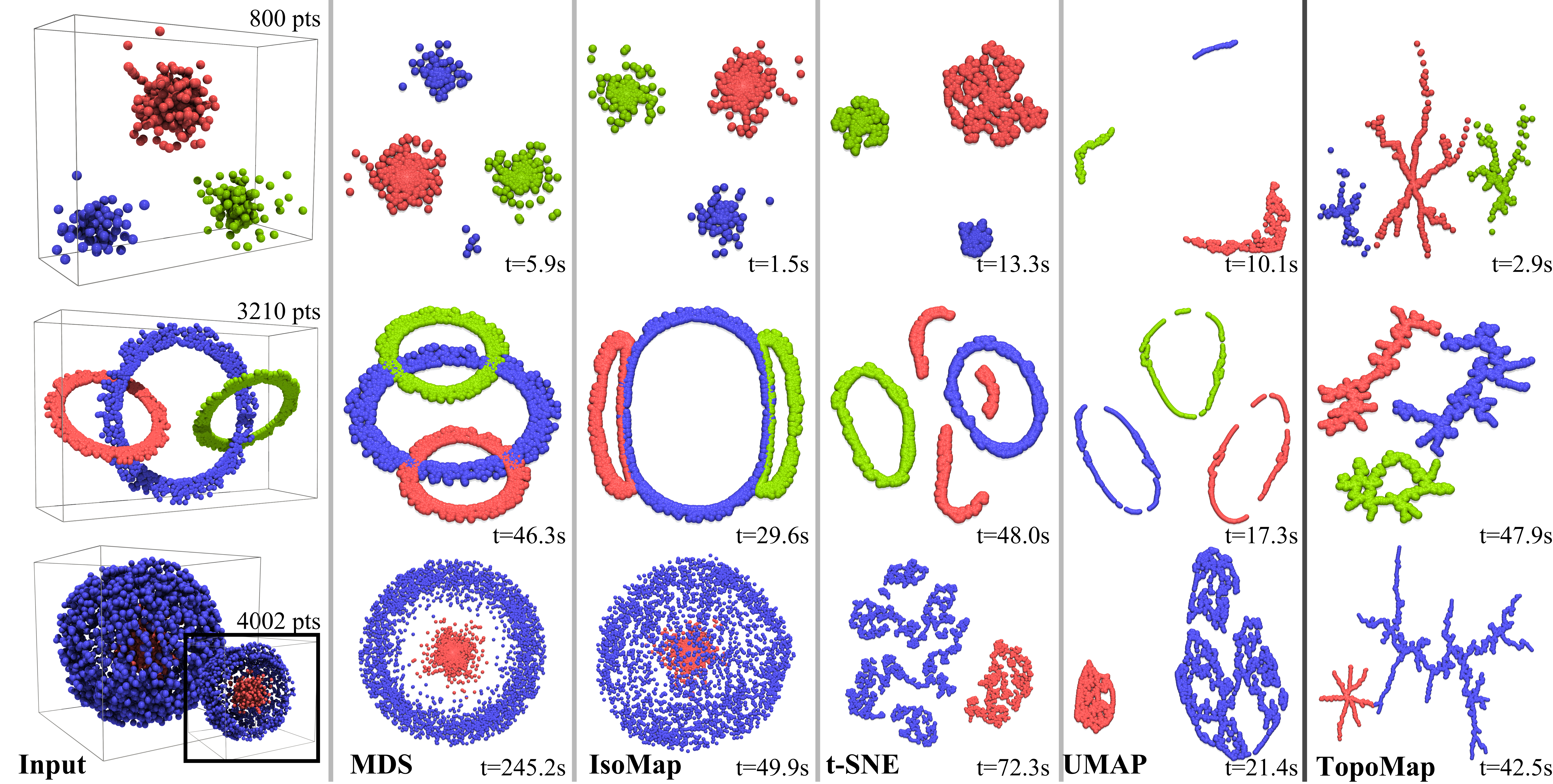}
 \caption{The result of mapping three dimensional data to a 2D space using geometry preserving projections: Classical MDS, Isomap, tSNE, UMAP; and the proposed topology preserving TopoMap method.  While geometry preserving methods tend 
either to split connected components or mix them up, TopoMap is guaranteed to
 preserve them, leveraging more reliable analysis.}
	\label{fig:teaser}
}

\vgtcinsertpkg

\begin{document}


\firstsection{Introduction}
\label{sec:intro}
\maketitle

Multidimensional Scaling (MDS) accounts for the problem of embedding data in a Cartesian space while preserving intrinsic properties of the data.
A particularly important task in the context of MDS is dimensionality reduction, which aims to map data from a $d$-dimensional 
to a $k$-dimensional Cartesian space where $k<<d$. 
In the context of visualization,
where the embedding space is 2D or 3D, MDS is typically called multidimensional projection (MDP). 

Over the last decades, a multitude of MDP methods have been developed to map high-dimensional data to a visual space
while preserving geometric properties such as the Euclidean distance between data objects. A main issue shared by
all those methods is that the preservation of geometric properties can only be guaranteed under very particular conditions.
Thus, errors and distortions are highly 
likely in the resulting mapping, 
introducing uncertainties to analytical procedures carried out from projection layouts~\cite{sacha2017visual}. 
For instance, structures observed in the point cloud resulting from a projection such as neighborhood relations might not be the ones existing in the original data,
thus potentially leading inexperienced practitioners to wrong conclusions.

Although a number of alternatives have been proposed to render the analysis of projection layouts more reliable~\cite{nonato:tvcg:2019}, 
few are focused on developing MDP methods with theoretical guarantees as to properties preserved by the mapping. 
Guaranteeing that a certain property is preserved exactly by the mapping makes the analytical process more reliable and meaningful,
ensuring that what is seen is indeed what takes place in the high-dimensional space.

This work introduces TopoMap, a novel MDP technique that is guaranteed to preserve topological structures during the dimensionality reduction process.
Specifically, TopoMap maps high-dimensional data to a visual space while preserving 0-homology (Betti-0) topological persistence 
as defined by a Rips filtration over the input data points. 
Intuitively, a Rips filtration grows a high dimensional ball around the data points, and adds an edge (or a high dimensional simplex)
to the filtration when two (or more) balls intersect.
In other words, the proposed method ensures that the topological filtrations over both the original as well as the projected data
generate the same connected components at the same instances of the respective filtrations.
The topological guarantee provided by TopoMap allows analysts to confidently explore high-dimensional data by visualizing 
which groups of objects are more tightly connected 
in the high-dimensional space.
As we show in the provided case studies, visualizing persistent components via Betti-0 preserving projections enables an intuitive 
analytical process, making the identification of objects with similar properties an easier task. Moreover, in contrast
to 
many distance-based (dissimilarity-based) projections, there is no uncertainty in 
the visual identification of groups (clusters) in the
layout produced by TopoMap. 

Besides enabling reliable mechanisms for data exploration, the proposed 
methodology can be used to assess and better 
understand distance-based projections.
Since TopoMap is guaranteed to preserve the connected components of a particular neighborhood graph structure, one can  
rely on it to analyze how those connected components are mapped by other projection methods. As a result, one can further understand
how distance-based MDP methods split or merge components, thus revealing regions with distortion. 

In summary, the main contributions of the work are:
\squishlist
\item A theoretical framework to support the design of a dimensionality reduction technique called TopoMap, which is guaranteed to preserve
the Betti-0 topological persistence defined by the Rips filtration over the data.
\item An optimization procedure that ensures the correct mapping of the connected components resulting from the filtration process.
\item An exhaustive evaluation using both labeled data as well as case studies over unlabeled data showing the potential of 
TopoMap to support the analysis of high-dimensional data as well as distance-based projection techniques.
\squishend
To the best of our knowledge, TopoMap is the first dimensionality reduction technique to provide guarantees as to the 
preservation of the topological properties of the Rips 
filtration of the data under analysis.

\section{Related Work}
\label{sec:rel-work}

In order to better contextualize the proposed methodology, we organize the related work in two main
parts, topological data analysis (which describes related 
work in topological data representations)
and topology-based multidimensional projection (which 
describes how these topological data representations can drive projection 
methods).

\subsection{Topological data analysis}
\label{sec_tda}

Topology-based methods \cite{edelsbrunner10} have been very 
popular in the last two decades to support advanced data analysis and 
visualization tasks \cite{heine16}.
By providing a 
concise, structural
representation of the data, these techniques greatly 
help in the visualization and analysis of the data.
They have been applied successfully to a 
variety of domains, such as 
astrophysics \cite{sousbie11, shivashankar2016felix},
biological imaging \cite{carr04, topoAngler, beiBrain18},
chemistry \cite{harshChemistry, chemistry_vis14, Malgorzata19}, 
fluid dynamics \cite{kasten_tvcg11},
material sciences \cite{gyulassy_vis07, gyulassy_vis15, soler_ldav19}, or
turbulent combustion \cite{laney_vis06, bremer_tvcg11, gyulassy_ev14}.

%

%

The Rips filtration~\cite{edelsbrunner10, ripser} is often used to
analyze the topology of high dimensional point clouds, and is motivated by the
work by Chazal and Oudot~\cite{DBLP:conf/compgeom/ChazalO08} who showed that 
Rips filtrations can provably capture the homology of the manifold sampled by the point cloud. 

Among the popular representations in topological data 
analysis, the \emph{Reeb graph} \cite{reeb1946points} is obtained by contracting 
to a single point each of the connected components of level sets of an input scalar 
field, resulting in a characteristic skeleton-like representation of the input 
data.
For discrete point cloud data, 
the \emph{Mapper} \cite{mapper} is an approximation of the Reeb graph of some 
user-defined function (often called \emph{lense} function) defined over a 
nearest neighbor graph of the input point cloud. 
%
Another popular abstraction is the Morse-Smale complex 
\cite{Defl15}, which is a cellular decomposition of the domain of an input 
scalar field, such that all the points of a given cell admit the same gradient 
integration extremities. 
For discrete point cloud 
data, the Morse-Smale complex has been used over the $k$-nearest neighbor 
graph of the input point cloud for clustering purposes 
\cite{DBLP:journals/jacm/ChazalGOS13}.
As discussed next, all of these representations
(Mapper, Reeb graph, Morse-Smale complex) have 
also been used as a driving data representation for dimensionality reduction.


Given the increasing popularity in using topology-based techniques for data analysis, 
it is not surprising that there are several open source tools and libraries available 
\cite{Nanda:Perseus, MariaBGY14, BauerKRW14, bubenik2017persistence, FasyKLM14, 
Morozov:Dionysus, AdamsTV14, ttk17}.



\subsection{Topology-based Multidimensional Projection}
\label{sec_topologyForReduction}

Multidimensional projection has long been a fundamental analytical tool, mainly
in the context of visualization~\cite{nonato:tvcg:2019, KrauseDFB16}.
In fact, the visualization community has not only proposed a number of MDP methods tailored to visual analytic tasks~\cite{joia:tvcg:2011}, but 
has also developed methodologies to facilitate the analysis of MDP distortions~\cite{aupetit:neuroc:2007,martins:candg:2014} 
and to enrich MDP layouts so as to uncover information hidden in the projections~\cite{gomez:tvcg:2016,joia:cgf:2015}. 
The extensive literature about MDS/MDP techniques has been organized over several books~\cite{borg97, dimensionReductionBook} and 
surveys~\cite{surveyDimensionReduction1,surveyDimensionReduction2,nonato:tvcg:2019}.
%
In order to emphasize our contribution, 
we focus only on techniques that explicitly rely 
on topological concepts to 
perform and assess multidimensional projections, disregarding distance preserving methods such as the classical MDS~\cite{dimensionReductionBook} and neighborhood preserving
techniques such as LLE~\cite{roweis2000nonlinear}, t-SNE~\cite{tsne}, 
and Lamp~\cite{joia:tvcg:2011}.
We refer interested readers to the 
above
books and surveys for a broader discussion about MDS/MDP methods.

\emph{Isomap}~\cite{tenenbaum2000global} is prossibly one of the first MDS techniques to resort to 
topological mechanisms to accomplish dimensionality reduction. Isomap aims to capture the topological (manifold) structure of the data 
through a graph representation from which geodesic distances
are computed. A number of variants of Isomap have been proposed, including Landmark (L-Isomap)~\cite{silva2003global}, out-of-sample~\cite{bengio2004out} and spatio-temporal extensions~\cite{jenkins2004spatio}.
An interesting variant of Isomap is the method proposed by Lee and Verleysen~\cite{lee2005nonlinear}, which tears a graph representation of the data so as to preserve 
essential (non-contractable) loops, thus enabling loop preserving manifold unfoldings. 
The recent work by Yan~et~al.~\cite{DBLP:journals/corr/abs-1806-08460} is 
another particularly interesting variant of Isomap (precisely, a variant of 
L-Isomap).
Similar to previous work on skeletonization \cite{kurlin_cgf15}, 
this approach identifies \emph{cycles} in the original data, but 
additionally aims at preserving these cycles when projecting the data 
to 2D. Specifically, it focuses
on 
the Mapper (cf. \autoref{sec_tda})
of a function 
defined on the KNN-graph structure of the data to select landmark points. 
The underlying idea is that the topology-based landmark selection captures the 
structure of the 1-dimensional homology groups of the data, 
which hopefully are preserved during the dimensionality reduction phase 
accomplished via regular L-Isomap. 
However, their approach does not take into account 0-homology groups which are 
therefore not preserved.
In contrast, the TopoMap method proposed in this work provides theoretical 
guarantees as to 0-dimensional homology group preservation, thus ensuring that 
the connected components visualized in the projection layout are 
 the same as in the original high-dimensional data, 
according to its Rips filtration.
Similar to 
Yan~et~al.~\cite{DBLP:journals/corr/abs-1806-08460}, 
Gerber~et~al.~\cite{gerber2010, gerber2013} introduced projection methods driven 
by topological data representations. In particular, it differs from our work in 
the sense that the introduced projections 
are driven by the network of cells of maximum dimension (called crystals) of 
the Morse-Smale complex (\autoref{sec_tda}). They do not aim at 
specifically preserving the persistence 
diagram of the Rips complex as studied in this paper and therefore encode a 
different information, specifically tailored for regression tasks.

%
%

In scientific visualization, 
Weber~et~al.~\cite{Weber:2007} introduced a terrain metaphor to 
provide an intuitive visualization of the topological features present in a 
volume scalar field. Due to occlusion, these features can be challenging 
to visualize when represented in their original 3D space. 
This work addresses this issue by constructing a 2D terrain whose elevation 
is carefully designed, such that the contour tree of the elevation map matches 
the contour tree of the original data in 3D. 
The resulting elevation can also be displayed as a planar heat map and the 
original data points can in principle be projected to this planar layout, by 
inserting each 3D point in the 2D region corresponding to its arc in the 
contour tree. 
This method can be interpreted as topology 
preserving, as the contour tree of the 2D heatmap  
is guaranteed by construction to be equal to the contour tree of the original data in 
3D. Note however, that the algorithm for constructing the terrain solely  
focuses on the contour tree and ignores the metric information coming from the 
original data. In particular, it places the root of the branch decomposition of 
the contour tree at the center of the layout and then arranges the children 
branches along a spiral trajectory \cite{Weber:2007}. This can have the effect 
of projecting in a small 2D neighborhood topological features which were 
originally arbitrarily far apart in 3D.
Harvey and Wang~\cite{HWTopo10} proposed algorithms to generate an 
ensemble of terrains each having the same contour tree as the input data. However, the
shortcomings described above apply to these terrains as well.

In a series of papers \cite{OesterlingHJS10, OesterlingSTHKEW10, 
OHJSH11, Oesterling0WS13}, Oesterling~et~al. extended this approach to the case 
of high-dimensional point clouds. This line of work is probably the most related 
to our approach. When extending the terrain metaphor to such data, the first 
difficulty is to derive a simplicial representation of the point cloud. In their
work,  Oesterling~et~al. suggest to use a specific adjacency graph called the 
Gabriel graph \cite{gabriel69}. The second challenge consist in deriving a 
scalar field on this graph which faithfully describes the data. The authors opt 
for a kernel density estimation of the point cloud (with a 
Gaussian kernel). From this point, the terrain metaphor \cite{Weber:2007} can 
be applied and the authors introduce various improvements 
\cite{OesterlingSTHKEW10, OHJSH11} based on contour profiles for instance 
\cite{Oesterling0WS13}.

In our work, by considering the Rips filtration, we focus our 
analysis on \emph{distances}, while Oesterling~et~al. focus on \emph{densities}. 
In that regard, these two approaches are complementary, just like distance-based 
and density-based clustering methods are complementary.
More importantly, the two approaches differ 
in the way the layout of the data in 2D is computed. 
As 
discussed above, the terrain metaphor \cite{Weber:2007} provides a constructive 
approach for computing the output 2D layout which discards the 
metric information of the original space, as acknowledged by the authors 
\cite{OesterlingHJS10}. Data points which are arbitrarily 
far in the original space can be projected arbitrarily close, and 
reciprocally. In contrast, our layout strategy enforces the preservation of the 
persistent homology of the Rips filtration. This enables to better take 
into account the metric properties of the data, and to some extent be more 
faithful to its original geometry. This tends to preserve the 
spatial relations between clusters (which are not taken into account in terrain 
metaphors). For instance, in \autoref{fig:teaser}, 
the central clusters in the data (top row: red, middle row: blue) are indeed 
projected in between the other clusters with our method (top 
and middle row, right column).

A subtle, yet important, distinction between our work and 
terrain metaphors 
\cite{OesterlingHJS10, OesterlingSTHKEW10, 
OHJSH11, Oesterling0WS13} is that our approach preserves topological 
features \emph{strictly} when projecting the data  to 
2D. In particular, the 0-dimensional persistence diagram of the Rips filtration 
of the projected data is strictly equal to that of the high dimensional data, 
by construction.
In contrast, terrain metaphors for high dimensional data
\cite{OesterlingHJS10, OesterlingSTHKEW10, 
OHJSH11, Oesterling0WS13} provide topology-preserving \emph{terrains}, but not 
necessarily topology-preserving \emph{projections}, as each data point is 
placed \emph{``at a random 
position along its} (density) \emph{contour''} \cite{OesterlingHJS10}.
Finally, note that to our knowledge, no public implementation of the terrain 
metaphors
is available.

The recently introduced UMAP approach 
\cite{2018arXivUMAP} is based on topological notions, namely category theory 
while our approach focuses on Persistent Homology~\cite{edelsbrunner10}. As 
reported by its authors, UMAP provides visual results which are highly similar 
to t-SNE. For this reason, it is often regarded as a faster, more 
modern and more scalable alternative to t-SNE, which still provides visually  
similar outputs.

Topological tools have also been the basis of methods designed to evaluate the quality of dimensionality reduction techniques. 
A good example is the work by Rieck and Leitte~\cite{rieck2015persistent}, 
which assesses the quality of a projection technique from
the 2nd Wasserstein distance between persistence diagrams computed from the 
original and projected data. In a follow up work, Rieck and 
Leitte~\cite{rieck2015agreement}
proposed the use of persistent homology to compare quality measures for dimensionality reduction, making possible to analyze the agreement of multiple quality measures,
thus identifying regions where different quality measures disagree 
the most.
Persistent homology has also been employed by Paul and Chalup~\cite{paul2017study} as a mechanism to validate dimensionality reduction methods when applied to particular benchmark data. 
As we shall prove later, TopoMap is guaranteed to preserve connected components under filtration, and is therefore
exact (no error) when comparing the 0-dimensional homology 
persistence diagrams generated by a filtration in the visual and original 
spaces respectively.

\begin{figure}
\centering
\includegraphics[width=\linewidth]{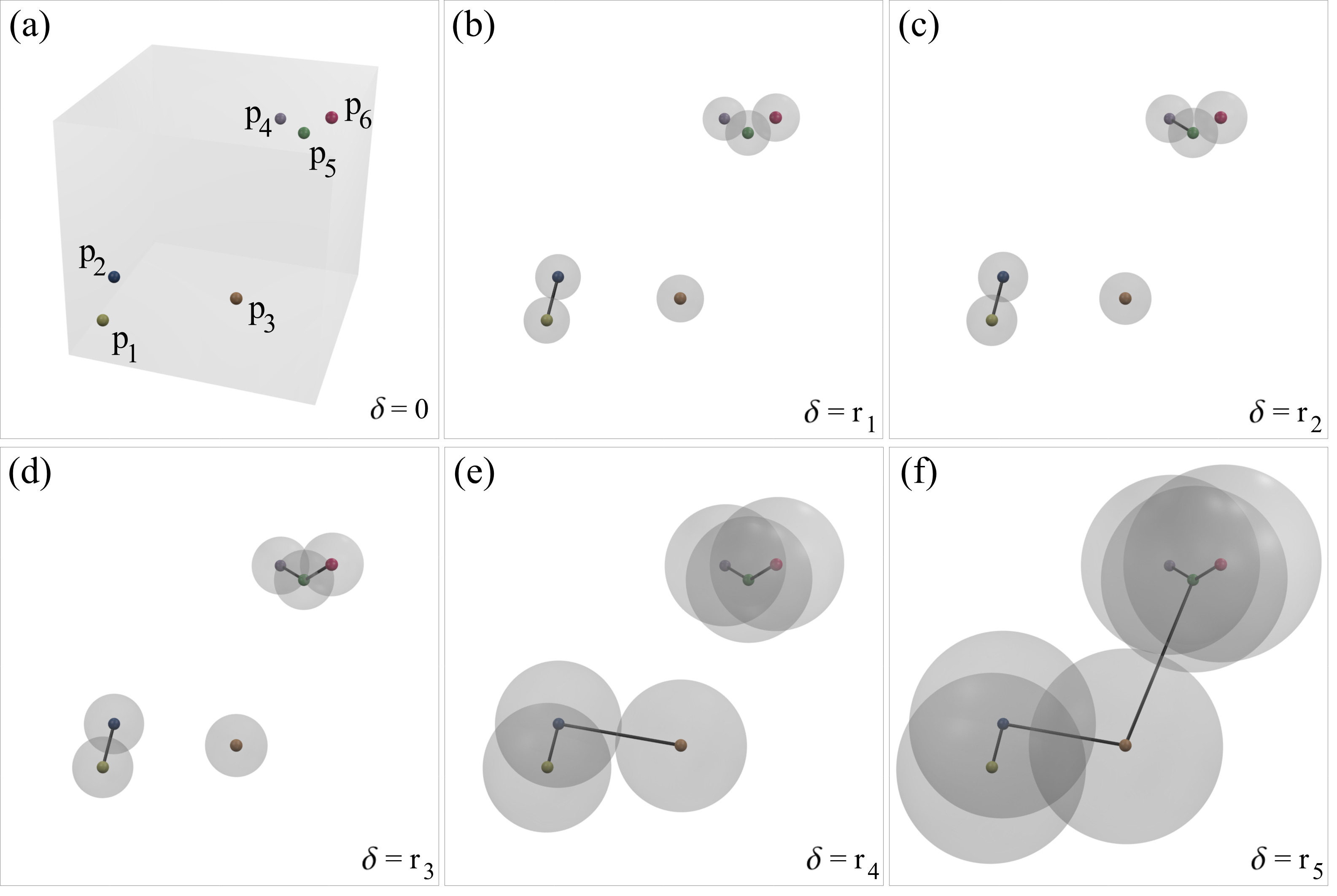}
\caption{The ball growth model used to analyze the topological properties of point data sets.
\textbf{(a)}~Input data.
\textbf{(b)--(f)}~Different stages of the filtration with increasing diameter $\delta$.
These stages correspond to the instant in the filtration when two components (0-cycles) merge
into one. The edge from the Rips filtration responsible for this merge is
also shown. Note that this collection of edges correspond to the minimum spanning tree
of the input points.
}
\label{fig:filtraiton}
\vspace{-0.2in}
\end{figure}

\section{Topology Preserving Projection}
\label{sec:protection}

Given a data set that is a collection of high-dimensional points in $\Rspace^d$,
a common topology-based approach to analyze this data is
to study the evolution of cycles in the simplicial complexes resulting from a 
Euclidean distance based Rips filtration over these points.
Our goal is to project the data onto $\Rspace^2$
such that above evolution for a subset of the cycles is
preserved in the projected space as well.

In this section, we first introduce the necessary notations and
formalize the problem that is of interest in this work.
We refer the reader to Edelsbrunner and Harer~\cite{edelsbrunner10} for a comprehensive
discussion on these topics.
Next, we describe a high level approach for solving the
problem, and discuss different choices that can be made
in the implementation of the high level solution.

\subsection{Problem Formulation}
\label{sec:formulation}

\myparagraph{Vietoris–Rips complex.}
Let $P = \{p_1,p_2,\ldots,p_n\}$ be a set of points in $\Rspace^d$.
Given a distance threshold $\delta$, the Vietoris–Rips complex~\cite{edelsbrunner10},
(or Rips complex), is defined as the set of all $k$-simplexes
$K \subseteq P$, $| K | = k+1, k \geq 0$,
such that $d(p_i,p_j) \leq \delta$,
$\forall p_i,p_j \in K$.
Here, $d(\cdot,\cdot)$ is the Euclidean distance.
Intuitively, the Rips complex for a distance threshold $\delta$
captures the shape of the data when each point $p_i$ is
replaced with a $d$-dimensional ball of diameter $\delta$ 
centered around it.
For example, consider the 6 points in $\Rspace^3$
shown in \autoref{fig:filtraiton}(a). 
\autoref{fig:filtraiton}(b)--\ref{fig:filtraiton}(f)
illustrates this shape for 5 different values
of the distance threshold $\delta$. 

\myparagraph{Rips Filtration.}
Consider a model where the distance threshold $\delta$
is increased from 0 to $\infty$. That is, the
$d$-dimensional balls are gradually grown in size.
A Rips filtration captures this growth model.

Consider an ordered set of simplexes $\Kspace_P = \{K_0 = \emptyset, K_1, K_2, \ldots, K_m\}$.
Let $\delta_i$, $i \in [0,m]$, be the smallest distance threshold
such that simplex $K_i$ is part of the Rips complex defined for $\delta_i$.
Then, the above ordered set is a Rips filtration if $\forall i,j$, $i < j$:
\begin{enumerate} \denselist
	\item $\exists\, l \leq i$ s.t. $K_i \bigcap K_j = K_l$; and
	\item $\delta_i \leq \delta_j$.
\end{enumerate}

\vspace{-0.1in}
\myparagraph{Topological Persistence and Persistence Diagram.}
Consider the growth as defined by the Rips filtration, wherein
the simplexes from the filtration are added one at a time.
That is, the $i^{\textrm{th}}$ iteration in this growth
will consist of the subset $S_i = \{K_0,K_1,\ldots,K_{i-1}\}$.
The addition of each new simplex can change the topology
of the underlying data, where the topology is captured by
the set of cycles in the simplicial complex defined by $S_i$.
More specifically, a new $k$-cycle, $k \geq 0$,
can either be created
or an existing $k$-cycle can be destroyed~\cite{edelsbrunner02}.
Informally, a $0$-cycle corresponds to a connected component,
a $1$-cycle to a loop, $2$-cycle to a void, and so on.
Given one such $k$-cycle, let $\delta_c$ be the threshold at
which this cycle is created, and $\delta_d$ the threshold at
which it is destroyed. Then the \textit{topological persistence}~\cite{edelsbrunner02}
of this $k$-cycle is defined as $\delta_d - \delta_c$,
and intuitively captures the lifetime of this cycle in
the given filtration.
Note that a cycle that is not destroyed has a persistence
equal to infinity.


The \textit{persistence diagram}~\cite{cohen-steiner05} plots all the cycles created during the
filtration as a scatter plot, where the coordinates of the point
corresponding to a cycle is its creation and destruction thresholds
(i.e., the $x$- and $y$-axes of this plot corresponds to
the creation and destruction thresholds).
%
%

\hidecomment{
An alternate visualization of the persistence of the cycles
is through the use of \textit{persistence barcodes}~\cite{Carlsson2004}.
\julienEdit{Each} cycle is represented as an \textit{interval} in $\Rspace$
between its birth and death threshold. \autoref{fig:persistence-vis}(b)
shows the barcode corresponding to the persistence diagram in
\autoref{fig:persistence-vis}(a).
}


\myparagraph{Problem Definition.}
Let $PD_P^k$ denote the persistence diagram restricted to $k$-cycles
computed using the Rips filtration over the point set $P$.
Given a set of points $P = \{p_1,p_2,\ldots,p_n\}$ in $\Rspace^d$, our goal is
to compute a corresponding set of points $P' = \{p_1',p_2',\ldots,p_n'\}$ in
$\Rspace^2$ such that $PD_P^0 = PD_{P'}^0$, where there is a one to one
correspondence between the connected components or 0-cycles (i.e., a point $p_i$
belongs to a 0-cycle w.r.t $P$ if, and only if, the point $p_i'$ will
belong to the corresponding 0-cycle w.r.t. $P'$).

In other words, the Rips filtration over the projected points $P'$
not only has the exact same connected components
during each iteration of the growth, but even the iterations
at which they are created and destroyed are the same
when compared to the Rips filtration over the high-dimensional points $P$.

\subsection{Approach}
\label{sec:algo}

%
Since we are interested only in the evolution of the set of connected
components, it is sufficient to consider only the
0- and 1-simplexes (vertices and edges respectively) of the filtration.
Consider the set of edges in the above filtration. Only a subset
of these edges result in a change in topology, or in other words,
merge two disconnected components into a single component.
The following lemma bounds the number of such \textit{topology changing edges}
in the filtration.

\begin{lemma}\label{lemma:topo-edges}
Given a Rips filtration defined over a set of $n$ points, there is exactly
$n-1$ topology changing edges that result in reducing the number of 0-cycles.
\end{lemma}
\begin{proof}
Consider an input with $n$ points. At the beginning of the filtration,
say at an infinitesimally small threshold $\epsilon > 0$,
there are a total of $n$ components each corresponding to an input point.
The addition of each topology changing edge reduces the count of connected components
by one. Thus, there exists exactly $n-1$ such edges until
there exists just a single connected component.
\end{proof}
The 5 edges in Figures~\ref{fig:filtraiton}(b)--\ref{fig:filtraiton}(f)
correspond to the topology changing edges in the Rips filtration
over the 6 points in \autoref{fig:filtraiton}(a).

Let $\Kspace_P^0 = \{\emptyset, p_1, p_2, \ldots, p_n, e_1, e_2, \ldots,
e_{n-1}\} \subset \Kspace_P$ be the subset of a filtration, where
$p_i$, $1 \leq i \leq n$, are the set of input points and $e_i$, $1 \leq i < n$,
corresponds to topology changing edges (in order of their appearance in
$\Kspace_P$). 
%
Note that we ignore all other edges in $\Kspace_P \setminus \Kspace_P^0$, since
they do not change the topology with respect to 0-cycles.
%

Consider only the ordered set of topology changing edges
$\Kspace_0 = \{e_1, e_2, \ldots, e_{n-1}\}$ from the
above filtration. By definition, the length of these edges satisfies $|e_1| < |e_2| < \ldots < |e_{n-1}|$.
While these inequalities might not hold in practice (two consecutive edges
could have the same length), a simulated small perturbation~\cite{EdelsbrunnerM90}
of the points can ensure this property holds.
The following lemma, which shows the equivalence between
$\Kspace_0$ and the Euclidean distance minimum spanning tree~(EMST) computed over $P$,
provides the basis for our projection algorithm.

\begin{lemma}\label{lemma:filtration-mst}
Given a set of points $P = \{p_1, p_2, \ldots, p_n\}$, let
$G$ be the complete weighted graph defined over $P$ such that
the weight of each edge $(p_i,p_j)$ is equal to the Euclidean distance
$d(p_i,p_j)$ between the corresponding end points.
Then, the ordered set of topology changing edges $\Kspace_0 = \{e_1, e_2, \ldots, e_{n-1}\}$
is precisely the set of edges of the minimum spanning tree~(MST), in increasing order of weight,
computed over $G$.
\end{lemma}
\begin{proof}
We prove this by showing, through induction, that the ordered set
of topology changing edges are the same as those added by the
Kruskal's algorithm~\cite{CLR01}.

Consider the first edge $e_1$ of the filtration. By definition,
it is the edge with the smallest length, and thus also the first edge
that is added by the Kruskal's algorithm.
Let, edges $e_1, e_2, \ldots, e_{i-1}$ be the first
$i-1$ edges added by the Kruskal's algorithm.
The induction hypothesis is that at this stage, the connected
components created by the filtration is exactly the same as the set of connected
trees created by the Kruskal's algorithm.

Now, consider the $i^{\textrm{th}}$ topology changing edge of the filtration $e_{i}$.
For sake of contradiction, say the $i^{\textrm{th}}$ edge
added by the Kruskal's algorithm is $e' \neq e_{i}$.
This implies that the edge $e'$ has length less than that of $e_{i}$,
and connects two connected subtrees together.
Then, by definition, $e'$ will occur before $e_{i}$ in the filtration,
and will also be a topology changing edge. Thus, the case
of $e' \neq e_{i}$ is not possible.
\end{proof}

The following proposition follows from the above lemmas and it guarantees
the existence of a mapping (projection) that retains the topology
with respect to 0-cycles in the projected space.
%



\begin{proposition}\label{lemma:topo-equivalence}
Let $P$ be a set of points in $\mathbb{R}^d$, 
$\Kspace_0=\{e_1,e_2,\ldots,e_{n-1}\}$ be the ordered subset of topology
changing edges in $\Kspace_P^0$ and
$C_P^i$ be the set of connected components
obtained during the filtration over $\Kspace_P^0$ after the addition of the first
$i$ topology changing edges $\{e_1,e_2,\ldots,e_{i}\}$. 
Let $\mathcal{M}:\mathbb{R}^d\rightarrow\mathbb{R}^k$ be a mapping that 
maps points in $P$ to $P'$, and $\Kspace_0'=\{e_1',e_2',\ldots,e_{n-1}'\}$
be the set of topology changing edges in $\Kspace^0_{P'}$.
Then, there exists at least one mapping $\mathcal{M}$ satisfying the following
properties:
\vspace{-0.1in}
\begin{enumerate}\denselist
\item[(a)] edge lengths $|e'_i| = |e_i|$, $\forall i \in [0,n-1]$;
\item[(b)] the components generated by the filtrations are identical, i.e., $C_{P'}^i = C_{P}^i$ $\forall i \in [0,n-1]$; and 
\item[(c)] $PD_{P'}^0 = PD_P^0$, where $PD_{P'}^0$ and $PD_P^0$ are the persistence diagrams of
$\Kspace^0_{P'}$ and $\Kspace^0_P$ respectively.
\end{enumerate}
\end{proposition}

We abuse notation in the above proposition when stating that $C_{P'}^i = C_{P}^i$.
What this notation means is that the mapping $\mathcal{M}$ establishes 
a one-to-one relation between the components 
in $C_{P}^i$ and $C_{P'}^i$, that is, every point in a component $C'\in C_{P'}^i$ is the image of a point
in the corresponding component $C\in C_{P}^i$.
Note that in the above proposition, guaranteeing properties (a) and (b) is a sufficient condition
for (c).

\setlength{\textfloatsep}{0pt}
\begin{procedure}[t]
	\small
	\caption{TopoMap()}
	\label{code:topomap}
	\begin{algorithmic}[1]
		\REQUIRE High dimensional points $P = \left\lbrace p_1, p_2, \ldots, p_n \right\rbrace$
		\STATE Compute the Euclidean minimum spanning tree $E_{mst}$ over $P$
		\STATE Let $E_{mst} = \{e_1,e_2,\ldots,e_{n-1}\}$ be the edges ordered on length
		\STATE Let $P' = \{p_1',p_2',\ldots,p_n'\}$, where $p_i' = (0,0)$, $\forall i$
		\STATE Let $C_i = \{p_i'\}$ be the initial set of components
		\FOR{each $i \in [1,n-1]$}
		\STATE Let $(p_a,p_b)$ be the end points of edge $e_i$
		\STATE Let $C_a$ be the component containing $p_a'$ and $C_b$ be the component containing $p_b'$
		\STATE \label{line:place-components}Place $C_a$ and $C_b$ in $\Rspace^2$ s.t. $\min_{p_j' \in C_a, p_k' \in C_b}\{d(p_j',p_k')\} = \textrm{length}(e_i)$
		\STATE Let $C' = C_a \bigcup C_b$
		\STATE Remove $C_a$ and $C_b$ from the set of components, and add $C'$ into this set
		\ENDFOR
		\RETURN $P'$
	\end{algorithmic}
\end{procedure}

The proof of Proposition 1 is constructive and is provided in 
\autoref{sec:placement}.
In fact, using the above lemmas, we design the iterative algorithm
shown in Procedure~\ref{code:topomap} that
projects a set of high-dimensional points onto $\Rspace^2$ while guaranteeing the properties stated in Propositon 1.
The algorithm places the points onto a plane such that
the minimum spanning tree edges are preserved.
In other words, it ``draws" the minimum spanning tree
maintaining the edge lengths.

The algorithm initially maintains all the points as a separate component,
and stores the minimum spanning tree edges as an ordered set.
In each iteration, the algorithm then adds the smallest edge from
this ordered set to connect two components, 
thus reducing the number of maintained components by one.
The length of the edge is preserved during this step,
that is, its placement is such that distance between the
two components is equal to the length of the connecting edge that, in turn, has the same 
length as its counterpart in the original space.
This edge is then removed from the edge set and the process repeated until
all edges are appropriately placed.
The key aspect of the algorithm is Line~\ref{line:place-components}
that places the points based on the minimum spanning tree edge lengths.
We describe different ways of accomplishing this in the next section.
The maintenance of the set of connected components is
accomplished using the union-find data structure~\cite{CLR01}.
\setlength{\textfloatsep}{\oldtextfloatsep}
\subsection{Building the Topology Preserving Mapping}
\label{sec:placement}

The TopoMap algorithm starts with $e_1$, the smallest topology changing edge in $\Kspace_0$. 
Placing the end points of $e_1$ (which are individual components at the start of this procedure) 
in a lower dimensional space such that this distance is preserved is straightforward. 
In other words, $e_1 = e_1'$.
Now, suppose that the first $i-1$ topology changing edges have been added while ensuring Proposition 1. 
In the $i^{th}$ step, let edge $e_i'$ be added so as to connect two components from $C_{P'}^{i-1}$ that are
counterparts of the components in $C_{P}^{i-1}$ linked by $e_i$. 
As mentioned in Line~\ref{line:place-components} of the algorithm, the goal now is to 
place these two components such that
the minimum distance between them is \textit{equal} to
$|e_i|$. If this condition is satisfied, then the properties 
$|e_j|=|e_j'|$ and $C_{P'}^{i}=C_{P}^{i}$, $\forall j\in\{1,\ldots,i\}$, are naturally attested.
By repeating the process for all $e_i$, $i\in\{1,\ldots,n-1\}$ we also guarantee that $PD_{P'}^0 = PD_P^0$. 
Therefore, proving Proposition~1 now requires showing that there exists a way to place two components 
connected by an edge $e_i'$ whose length is $|e_j|$.
In fact, there are several ways in which this can be accomplished
as we show next.
Note that, there \textit{always} exists a valid solution to
this problem.

\begin{figure}[t]
	\centering
	\includegraphics[width=\linewidth]{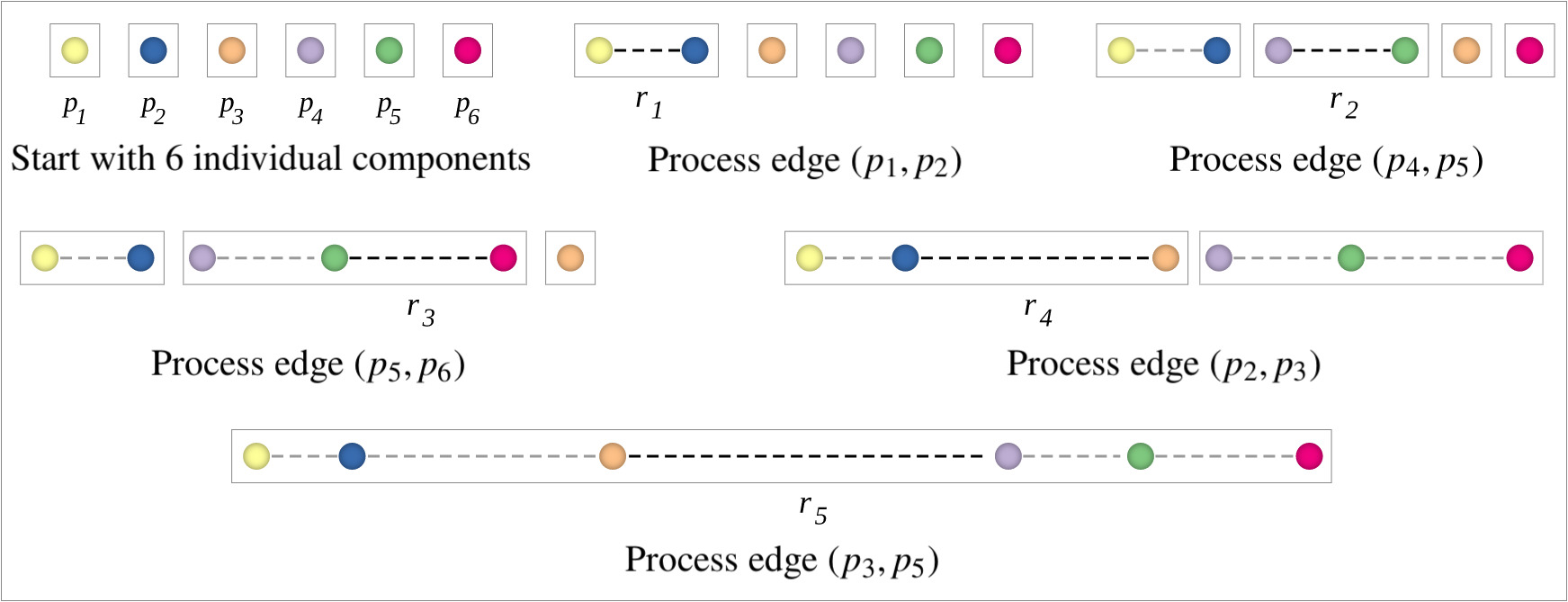}
	\vspace{-0.2in}
	\caption{Projecting the points from \autoref{fig:filtraiton} in 
		1-dimensional space.
		Each iteration processes one edge (in increasing order to length) from the minimum spanning tree.}
	\label{fig:trivial}
	\vspace{-0.25in}
\end{figure}

\myparagraph{A solution in 1-dimensional space.}
Consider two sets $C_1$ and $C_2$, where each point in this
set is associated with a $x$ and $y$ value corresponding
to its 2D coordinates.
Let $p_r \in C_1 | p_r.x > p'\!.x, \forall p' \neq p_r \in C_1$.
In other words, $p_r$ is the \textit{rightmost} point in $C_1$.
Similarly, let $p_l \in C_2 | p_l.x < p'\!.x, \forall p' \neq p_l \in C_2$
be the \textit{leftmost} point in $C_2$. Let the input edge length
be $d$.
The trivial solution is to simply translate all points in $C_1$ such that
$(p_r.x,p_r.y) = (\frac{-d}{2},0)$ and $(p_l.x,p_l.y) = (\frac{+d}{2},0)$.
\autoref{fig:trivial} illustrates this procedure for the
example shown in \autoref{fig:filtraiton}.

\begin{figure}[t]
	\centering
	\includegraphics[width=\linewidth]{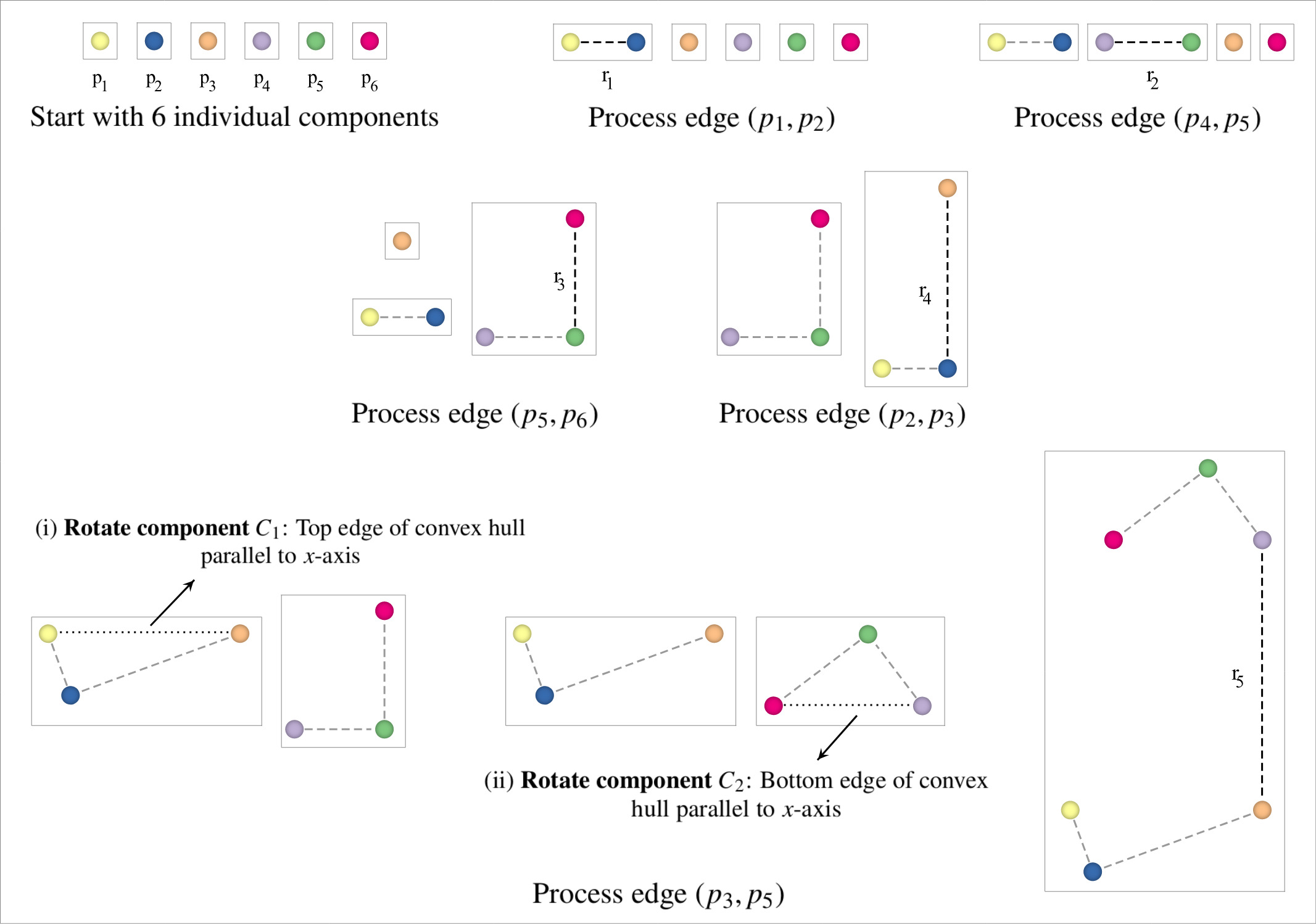}
	\vspace{-0.2in}
	\caption{Projecting the points from \autoref{fig:filtraiton} in 
		2-dimensional space.}
	\label{fig:deterministic}
	\vspace{-0.2in}
\end{figure}

\myparagraph{A geometric 2D solution.}
Note that the above solution places all the points
only along the $x$-axis. Thus, to obtain a more compact solution
that also uses the second dimension, we modify the
above solution as follows, by arranging components in the 
plane with local rotations, similarly to circular layout strategies in tree 
drawing \cite{Rusu2013TreeDA}.
Let $hull(C_1)$ and $hull(C_2)$ be the convex hulls of the
two components. Pick an edge $e_t$ from $hull(C_1)$ and $e_b$
from $hull(C_2)$. Transform (rotate) $C_1$ such that $e_t$
is parallel to the $x$-axis, and is the topmost edge of the
convex hull (i.e., has the highest $y$- coordinate).
Similarly, transform $C_2$ such that $e_b$ is also
parallel to the $x$-axis, but is the bottommost edge.
Let $left(e)$ denote the left endpoint of the edge $e$.
Now, translate component $C_1$ such that $left(e_t) = (0,0)$,
and component $C_2$ such that $left(e_b) = (0,d)$.
Alternatively, the right endpoints of the edges $e_t$ and $e_b$
can be used as well to align the two components.

There are different ways in which the edge of the convex hull
can be selected (as well as to decide which end point is used for the alignment).
We decided to choose the edge that
contains one of the end points of the minimum spanning tree edge
that is under consideration.
In case this is not possible, we choose the edge closest to
this point. The intuition here is to not only preserve the
connected components after every iteration, but to also
try and preserve the end points of the minimum spanning tree edges
as much as possible.

\autoref{fig:deterministic} illustrates this procedure for the
example points in \autoref{fig:filtraiton}. Note that the
addition of the first two edges result in the same
state as in the 1D solution above. However, when the third edge
$(p_5,p_6)$ is added, then the point $p_6$ is placed 
in a perpendicular orientation.
When the last edge $(p_3,p_5)$ is processed, since both components
have more than two points each, the convex hull is used to
perform the alignment by appropriately transforming both components.

\myparagraph{An optimization-based 2D solution.}
During data analysis, in addition to preserving the topology changing edges of
the filtration, it might be beneficial to also possibly preserve other
properties as much as possible. In this section, we show how our projection
approach can be tuned to support such modifications.

For example, it might be natural to consider a case where we are also interested
in keeping the resulting projection ``compact", in the sense that we want the
points to be as close to each other as possible, while still ensuring that the
distance between the two components, $C_1$ and $C_2$, is equal to the given
value. One way of doing this is to minimize the sum of squared distances between
the points in $C_1$ and $C_2$ after the placement. This can be achieved using
the optimization model described next.


First, fix one of the components, say $C_1$, and expand it to contain all points
in the plane that have distance to $C_1$ less or equal $d$ (this is the region
which \textit{should not contain} any point from $C_2$). This is achieved by
considering lines that are parallel and at a distance $d$ to the edges of
$hull(C_1)$. Since this expanded set of lines is also a convex hull, its inner
region can be described by a set on linear inequalities on the plane. Let this
set of linear inequalities be denoted by $A\textbf{x} \leq b$, $A \in
\mathbb{R}^{k \times 2}, b \in \mathbb{R}^k$.

\begin{figure}[t]
	\centering
	\includegraphics[width=\linewidth]{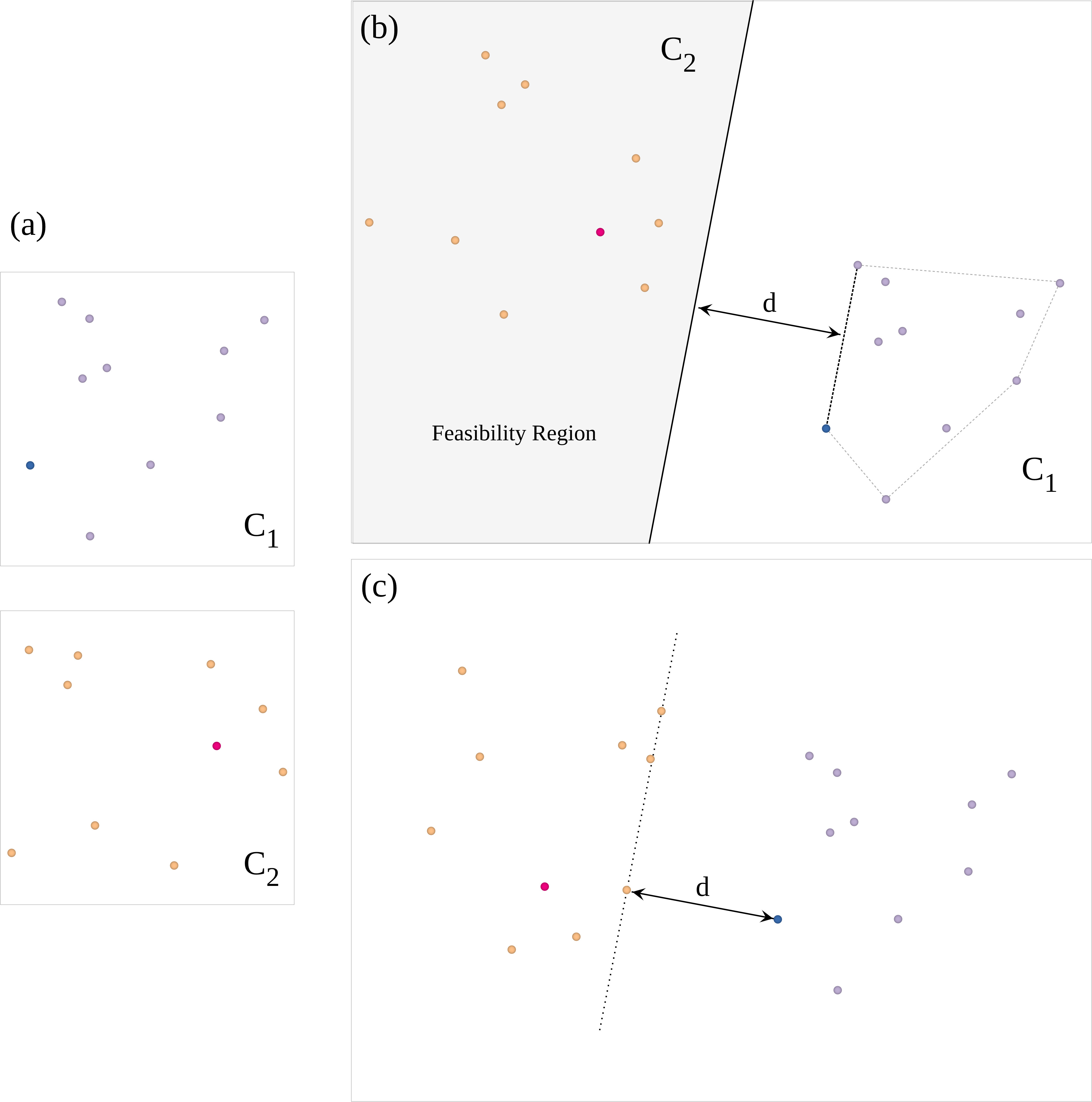}
	\vspace{-0.2in}
	\caption{Solving the optimization model to place two sets of points.
		\textbf{(a)}~Two components that are to be merged.
		\textbf{(b)}~The feasibility region with respect to the highlighted edge of $hull(C_1)$
		is shaded gray. $C_2$ is initially placed in this region and the optimization is solved.
		\textbf{(c)}~Solution.
	}
	\label{fig:optimization}
	\vspace{-0.2in}
\end{figure}

The goal then becomes to find the rigid motion (rotation plus translation) that
applied to $C_2$ minimizes the sum of the squared distances to the expanded convex
hull without penetrating it.
Formally, this problem can be mathematically formulated as:
\vspace{-0.1in}
\begin{align*}
\min_{\theta, t} &\quad \sum_{p_1 \in C_1, p_2 \in C_2} \|p_1 - (R(\theta) p_2 + t) \|^2 \\
\text{s.t.} &\quad A(R(\theta)p + t) \text{ is not strictly smaller than } b, \forall p \in C_2.
\vspace{-0.1in}
\end{align*}
where $\theta$ represents an angle with rotation matrix $R(\theta)$ and $t$ is a
translation vector. Note that it is possible to consider only the points that
define the convex hull of $C_2$ above.

This problem can be cast as a mixed integer nonlinear optimization problem.
The integer variables are needed because the constraints in the above model are
actually a ``union of sets" instead of an ``intersection of the sets" that is
usual in optimization. Unfortunately, solving such problems for a 
large
number of points is impractical. We therefore decided to use a simplified
heuristic that (i)~individually optimizes with respect to each edge of the
convex hull $hull(C_1)$; and (ii)~minimizes the sum of distances between points
in $C_2$ to a single point in  $C_1$. Formally, let $A_i$ be a row of the matrix
$A$ and $b_i$ the respective right-hand side.
Let $p' \in C_1$ be a point of interest. We first solve the following
optimization problem for $i = 1, \ldots, k$.
\vspace{-0.1in}
\begin{align*}
\min_{\theta, t} \quad \sum_{p \in C_2} \| p' - (R(\theta) p + t) \|^2 \text{~~~}
\text{s.t.} \quad A_i(R(\theta)p + t) \geq  b_i, \forall p \in C_2.
\vspace{-0.1in}
\end{align*}
We then consider as final solution the one that obtained the smallest objective
value.
If we also want to preserve the edge from the filtration in the
projection, weights can be applied to the objective function above, such that
the edge endpoint in $C_2$ has higher weight when compared to other points.
In our implementation, we consider $p'$ to be the end point of the edge that is
being processed in that iteration.

\autoref{fig:optimization} illustrates this optimization process. It 
shows
sets of points $C_1$ (colored violet) and $C_2$ (colored orange) that are to be
placed at a distance $d$ from each other 
(\autoref{fig:optimization}(a)). The
points colored red and blue correspond to the filtration edge under
consideration.
The blue point is chosen as the point of interest in order to minimize the
objective function of the optimization. The points in $C_2$ are first randomly
placed in the feasibility region corresponding to one of the edges of
$hull(C_1)$, and the optimization problem is solved
(\autoref{fig:optimization}(b)). The resulting solution is shown in
\autoref{fig:optimization}(c).

While the simplified optimization model is also nonlinear (due to the rotation),
it does not have integer variables and can then be solved by standard nonlinear
optimization algorithms. 
Note that the final solution is not guaranteed to have two points, one in $C_1$
and the other in $C_2$, at exact distance $d$ (and therefore, do not
satisfy the required filtration constraint). However, this can still be ensured
by sliding $C_2$ parallel to the edge of $hull(C_1)$ that is associated to the
solution obtained in the optimization process.

\myparagraph{Implementation.}
TopoMap was implemented using C++. It can be divided into
two phases.
First is to compute the Euclidean distance minimum spanning tree,
for which we used the implementation of the dual tree EMST algorithm~\cite{March2010}
provided by the \textit{mlpack} library~\cite{mlpack2018},
and has a time complexity of ($O(N\log N \alpha(N))$, 
where $N$ is the size of the input.
The next phase is to layout the points. Each iteration of TopoMap
aligns two components corresponding to the MST edge being processed.
We use the union-find data structure to maintain the list of components,
which can be accomplished in $O(N\alpha (N))$ time.
The convex hull of the resulting merged component is then computed using the qhull library~\cite{qhull},
which takes $O(n\log n)$ time to compute the convex hull of $n$ points.
However, since in each iteration, we use only the points in the convex hull
of the individual components, $n << N$ in practice. On several large data sets, we found that
the layout phase using the geometric approach scaled linearly with the
input. On the other hand, computing the EMST became the primary bottleneck
when 
increasing dimensions.

For the optimization based approach, we use the Algencan~\cite{algencan1,algencan2} library for solving our
optimization model. It is a robust and
high performance implementation of the augmented Lagrangian method for nonlinear
optimization problems whose code is freely available.
As expected, the optimization approach was slower than the geometric approach. 
However, the main bottleneck was still the EMST phase 
especially for large point clouds.

\subsection{Robustness}
\label{sec:robustness}

From the topological perspective, it is well known that the persistence diagram 
is 
robust to noise~\cite{cohen-steiner05}, especially in the 
context of topology inference~\cite{DBLP:conf/compgeom/ChazalO08}: small 
displacements of points in the original space induce small variations in the 
persistence diagram.
Since TopoMap strictly preserves
the
persistence diagram, topological robustness to noise is guaranteed by definition.
On the other hand, the ordering
of the filtration might change slightly due to the noise induced perturbation. 
Thus, with respect to the actual projection itself, the locations of the
points in 2D space might vary marginally when using the geometric solution. 
Note that the optimization-based approach, being 
non-deterministic, 
can produce different layouts even for the same input when run multiple times.
However, since the 
connected components (that represent the points in the persistence diagram)
are robust, these components are \textit{always} maintained by the projection.

\begin{table}
	\centering
	\small
	{
		\caption{Data sets used in our experiments.}
		\label{tab:data}
		\vspace{-0.1cm}
		\begin{tabular}{|l|c|c|c|}\hline
			\textbf{\Dataset} & \textbf{\# Instances} & \textbf{\# Attributes} & \textbf{\# Classes} \\ \hline
			Iris~\cite{uci}  & 150 & 5 & 3 \\ \hline
			Seeds~\cite{uci}  & 210 & 8 & 3 \\ \hline
			Heart~\cite{uci}  & 261 & 11 & 2 \\ \hline
			Cancer~\cite{uci}  & 699 & 11 & 2 \\ \hline
			Mfeat~\cite{uci}  & 2000 & 64 & 10 \\ \hline
			MNIST~\cite{mnist}  & 20000 & 784 & 10 \\ \hline
			Urban  & 17520 & 6 & not labeled \\ \hline
		\end{tabular}
	}
	\vspace{-0.6cm}
\end{table}

\begin{figure*}
	\centering
	\includegraphics[width=0.8\linewidth]{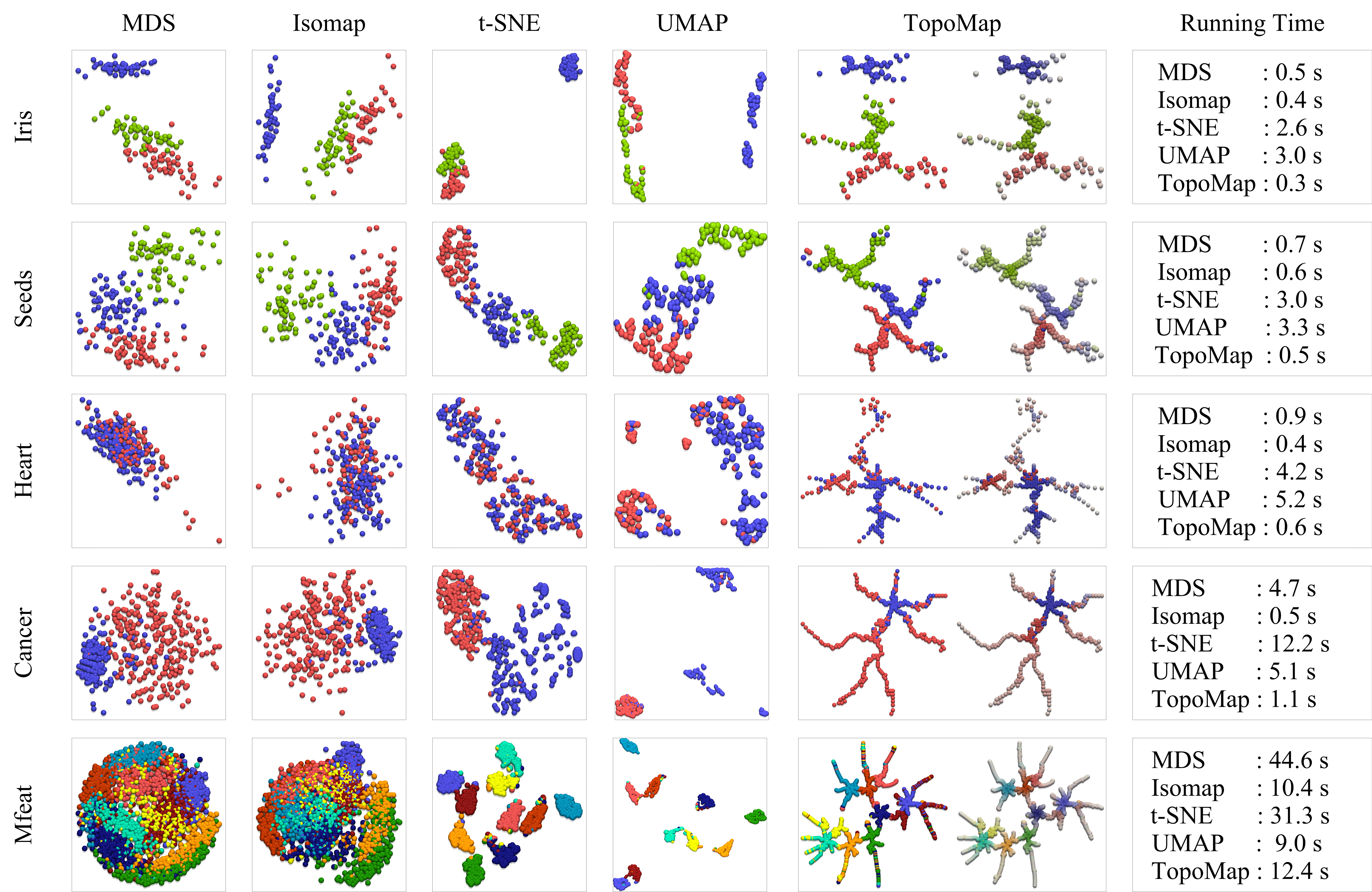}
	\vspace{-0.1in}
	\caption{Layouts produced by MDS, Isomap, t-SNE, UMAP and TopoMap when applied to the five first \datasets in Table~\ref{tab:data}. Right
		images in the TopoMap column highlight in colors the denser areas in the left images.}
	\label{fig:compare}
 	\vspace{-0.2in}
\end{figure*}

\section{Mapping Evaluation and Interpretation}
\label{sec:eval}

In this section we present the results of applying TopoMap to project high-dimensional data to a visual space ($\Rspace^2$). 
Our goal in this section is to analyze the properties of the layout produced by TopoMap, the way it visually encodes the information contained
in the high-dimensional data, and how much ``readable" the TopoMap layout is when compared to the ones produced 
by dissimilarity preserving projection methods. 

To facilitate the above analyses, we use several \datasets (see Table~\ref{tab:data}) 
having different numbers of instances and dimensions,
some of which are labeled (i.e., the classes are known for the instances).
We used the implementations provided by scikit-learn~(v0.19.0) for existing methods. 
All experiments were run on a machine with an Intel(R) Xeon(R) CPU E5-2630 v2 running at 2.60GHz
and 
64 GB of 
memory.



\hidecomment{
\subsection{Checking Theoretical Properties}
\gustavo{In this subsection we should starting by showing that persistent diagrams computed in the projected and original spaces are really identical.
We should be very direct here, if possible showing an example in 3D, showing that the connected components up to a certain filtration level are the same.}
\harish{This is theoretically proved. I still don't see the point of having this. All it accomplishes is that our code doesn't have bugs!}
}

\subsection{Layout Interpretation}
\label{sec:layoutinterp}

\autoref{fig:teaser} compares the layout produced by MDS, Isomap, t-SNE, UMAP, 
and TopoMap for 3 synthetic \datasets with well known properties.
The first is simply a set of points sampled from three Gaussians, the second is points sampled from three rings,
while the third is sampled from two concentric spheres. This figure 
illustrates the ability of TopoMap to preserve the connected components observed 
in the original space and to nicely reflect their relative adjacencies.
\autoref{fig:compare} performs the same comparison with respect to the first 
five \datasets in Table~\ref{tab:data}.
%
One can notice from these examples that the layouts resulting from TopoMap are quite different from
the ones produced by dissimilarity-based methods. 
This is not a surprise, since TopoMap preserves $n-1$ distances between connected components while
dissimilarity-based methods try to preserve $n^2$ distances (or distributions in the case of t-SNE) between instances.

%

\myparagraph{Star Shaped Ensembles.}
TopoMap produces a layout made up of star shaped ensembles with branches connecting and emanating from them. 
One way of interpreting this layout is through the 
use of the equivalence between 
the 0-homology filtration of the Rips complex and 
hierarchical clustering using the single-linkage 
criterion~\cite{carlsson09}. 
The connected components built during the filtration \textit{are exactly the 
same} 
as the clusters formed when 
moving up in the hierarchy. In other words, hierarchical 
clustering with single-linkage produces, by construction, identical results 
when considering the 
input (high-dimensional) data and the two-dimensional projection provided by 
TopoMap.
Thus, when interpreting our projections, users should visually 
identify centers of stars, as these correspond to clusters in the data (these 
also tend to correspond to the denser parts of the projection, see the 
TopoMap results of \autoref{fig:compare}, right column). 
On the contrary, the tips of the stars' branches should be interpreted as 
outliers or points lying at the boundary between clusters (these correspond to 
the least dense regions of the projections). Notice from \autoref{fig:compare} 
that there is a good correspondence 
between the star ensembles and the classes of data instances. 
If using the star shaped ensemble to guide the exploration, TopoMap enables visual analysis 
that does not demand a great cognitive effort to figure out which are the main groups of instances in the data.

Overall, compared to dissimilarity-based methods, TopoMap is equally, if not more, informative. 
In fact, except for the Heart \dataset, 
one can easily build a visual correspondence between star shaped ensembles and classes. 
However, even in the Heart, TopoMap
indicates the presence of groups of similar instances while the 
layouts resulting from MDS, Isomap and t-SNE are meaningless.
In the Cancer \dataset, MDS and Isomap clearly reveal one well defined group (blue dots), 
however, without the labels, it would be difficult  
to claim that the red points make up a class. The same is true with t-SNE as well, 
which clearly pinpoints the compact class (in red),
but it splits the blue class into a number of local clusters, 
increasing the potential for misleading interpretation. 
TopoMap, on the other hand, shows two star shaped ensembles, one well defined
and another more elongated, indicating the presence of two classes, one of them not so compact. 

\myparagraph{Density and Dispersion.}
The right most images in \autoref{fig:compare} (TopoMap column) 
highlight in colors the denser regions in each layout 
produced by TopoMap, while gray regions correspond to less dense areas. 
In particular, we use a Kernel Density Estimator (KDE) with a 
Gaussian kernel (one Gaussian is centered at each point in 2D and the sum of 
the contributions of all Gaussians is considered as a density estimation at 
each point). We additionally use an opacity transfer function, driven by this 
density estimation, that the users can further adjust if needed (by default, 
a simple threshold at half of the maximum 
estimated density).
%
%
%
The density-based  visualization makes it easier to identify tightly connected groups. 
Although density-based visualizations have been used to evaluate dissimilarity-based methods~\cite{martins:candg:2014},
the presence of errors and distortions prevent the analysis from being accomplished with high confidence~\cite{aupetit:neuroc:2007}. 
Note that the ``centers" of the starred ensembles 
correspond to denser areas of the layout, thus corresponding to tightly grouped data instances.
This is evident in the examples involving the Heart and Cancer datasets, 
where a density analysis in the layout resulting from MDS, Isomap, and t-SNE
would be of little use.

\myparagraph{Branches and Outliers.}
When considering the above mentioned density based visualization,
it is easy to see that branches stemming from the starred ensembles 
typically are low density regions.

These branches are essentially of two types: those connecting the starred ensembles;
and the ones emanating outwards from the stars. 
The latter 
is composed of points whose neighborhoods are not tightly connected.
From the hierarchical clustering perspective, these can be considered 
as single point clusters (outliers for example)
which merge with already existing large clusters as one moves up the hierarchy.

For example, the TopoMap projection of the Cancer dataset in 
\autoref{fig:compare}
contains one compact and another more sparse class. The sparse class (red) gives rise to a starred ensemble with a small "center" 
and long branches emanating from it. The density visualization, coupled with the guarantee that the 
topology changing edges' lengths are preserved by TopoMap, gives us confidence
to claim that the longer branches comes from the sparser class. Moreover, 
outliers tend also to be part of the loose branches,
mainly in less dense areas of the layout. This fact can be observed in the projection of Mfeat dataset, where
classes become mixed in loose branches (TopoMap column left image), but not in 
the center of the ensembles.

\begin{figure}[t]
\centering
\includegraphics[width=\linewidth]{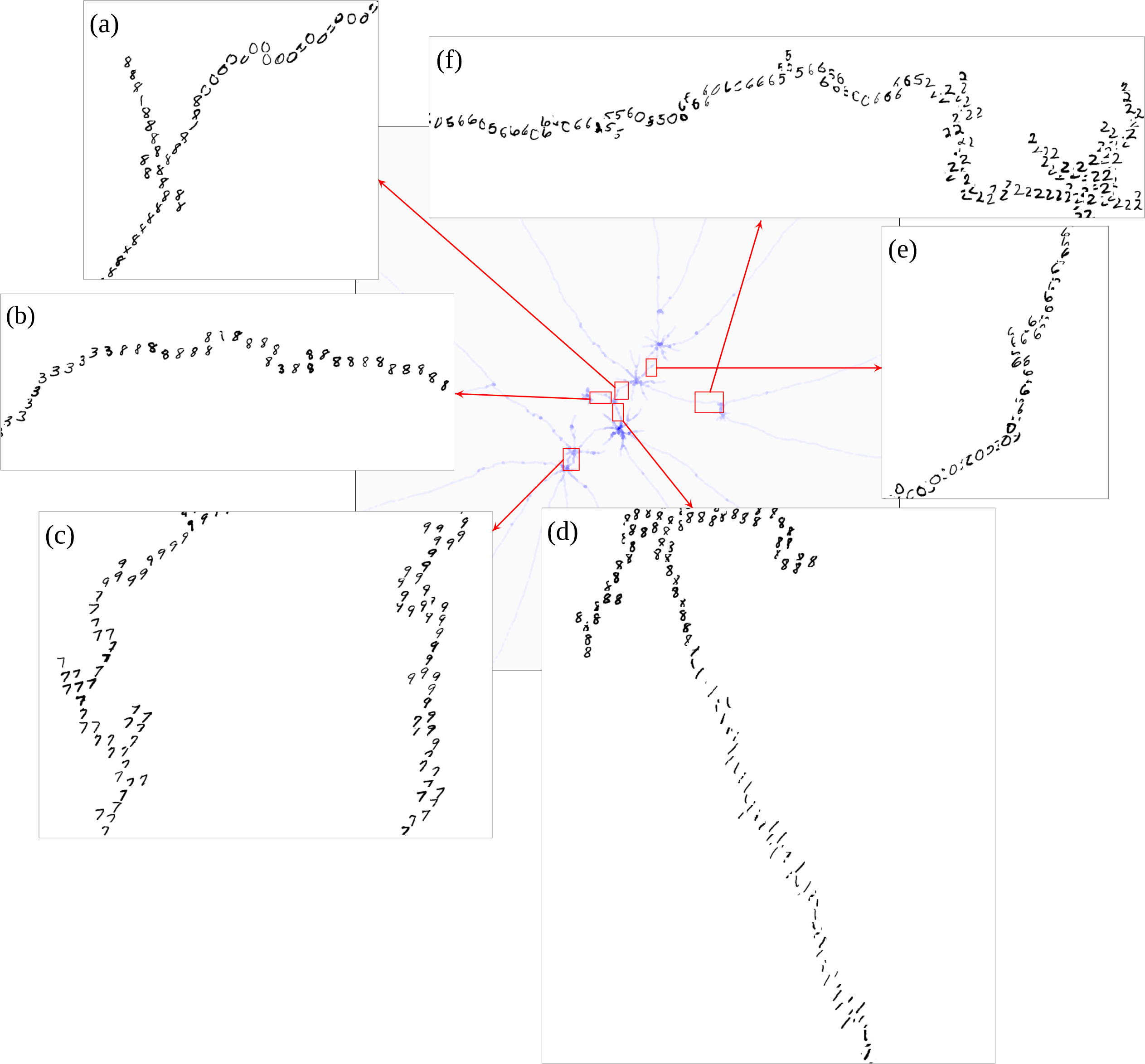}
\vspace{-0.2in}
\caption{Mnist data projected using TopoMap (using cosine distance). Transitions between the different starred ensembles clusters:
	(a)~0 and 8.
	(b)~3 and 8.
	(c)~7 and 9.
	(d)~1 and 8.
	(e)~0 and 6.
	(f)~class 2 while being a cluster, is far from 0 and is connected to it via outliers.
}
\label{fig:mnist}
\vspace{-0.25in}
\end{figure}

\begin{figure*}[t]
	\centering
	\includegraphics[width=0.9\linewidth]{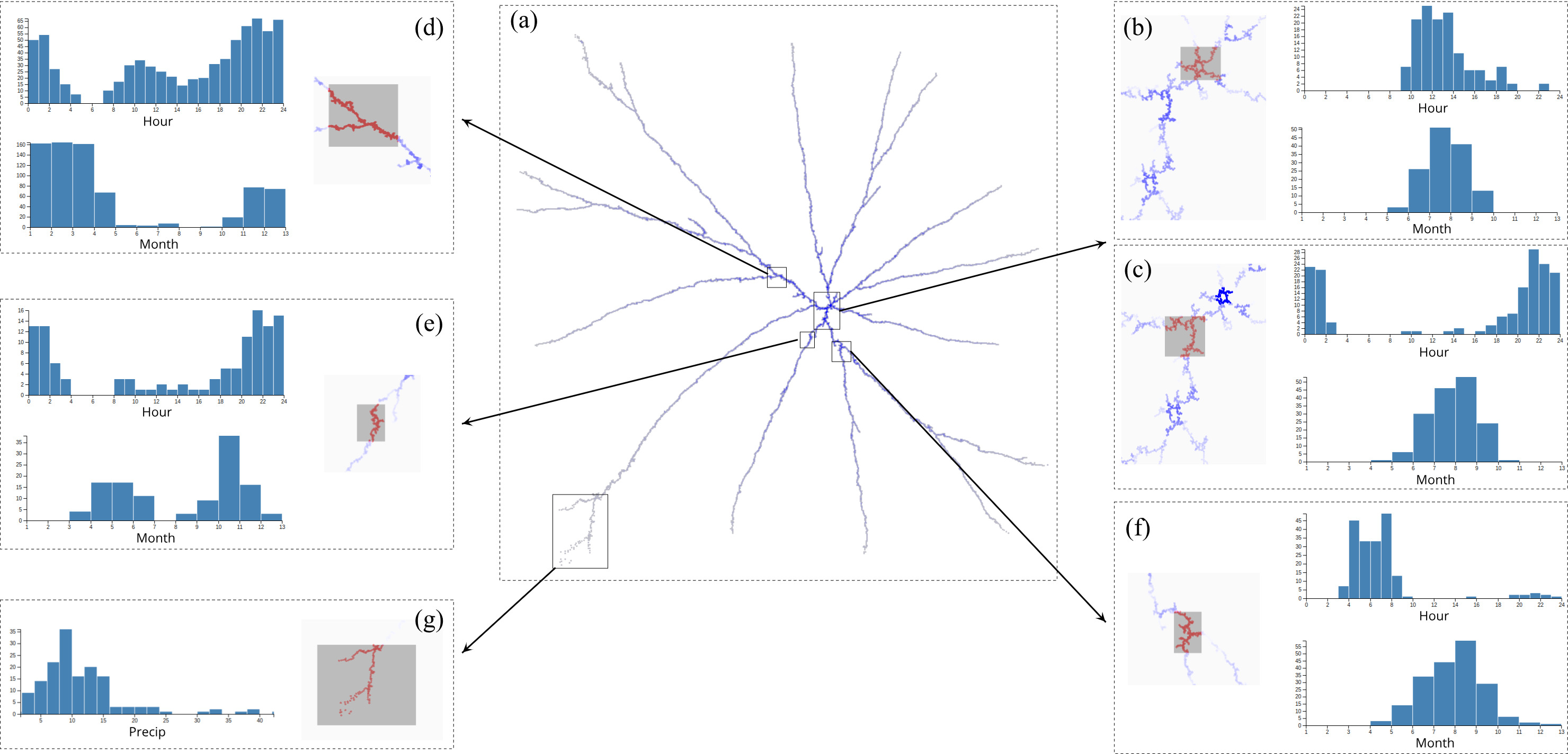}
	\vspace{-0.4cm}
	\caption{Analyzing Times Square using TopoMap. 
		\textbf{(a)}~Projection obtained using TopoMap.
		\textbf{(b)--(g):}~Different clusters are selected and the temporal 
distributions of the selected points visualized as a histogram. 
	}
	\label{fig:times-square}
	\vspace{-0.2in}
\end{figure*}

\myparagraph{Transitioning Between Clusters.} 
Branches connecting the centers of starred ensembles tend to encapsulate 
instances that lie between clusters, and 
represent a transition from one group to another. 
To illustrate this property, we used a uniform sample of 20000 instances from the MNIST data set, 
and projected it using the angular distance as distance metric (see 
\autoref{sec:discussions} for a discussion on this).
\autoref{fig:mnist} highlights the different transitions between the 
star ensembles captured by TopoMap.
In particular, note that the cluster corresponding to class \textbf{8} transitions to \textbf{0}~(a), \textbf{3}~(b), 
as well as \textbf{1}~(d). Other transitions such as 
from class \textbf{7} to \textbf{9} (c) and from \textbf{6} to \textbf{0} (e) can also 
be clearly seen. When well defined clusters are far apart from each other,
as is the case of cluster corresponding to class \textbf{2} 
(\autoref{fig:mnist}(f)), 
the branches emanating from are seen to be formed by ``outliers" lying in between 
the clusters. 
Notice, however, that the TopoMap layout clearly shows clusters located 
far apart in the layout, making it easy for users
to be aware of which branches are more prone to be made up of outliers 
(using the density based visualization to help this process).

In general, dissimilarity-based techniques capable of emphasizing clusters such as t-SNE and UMAP does not capture well
the transitioning between the clusters. In contrast, techniques capable of grasping transitions, such as Isomap, do not emphasize 
clusters well. Therefore, besides its theoretical guarantees, TopoMap bears properties difficult to be simultaneously present in 
dissimilarity based projection methods.

\myparagraph{Information loss.}
There are two main scenarios that can result in a loss of information during the TopoMap projection.
First, since the focus is on preserving the 0-cycles, any information with 
respect to higher dimensional cycles is lost.
The 3 rings example in \autoref{fig:teaser} is one such instance, where the 
1-cycles formed by 
the 3 main components
are simply represented as 3 star ensembles. A similar loss can be seen in the 
concentric spheres example, where the 
2-cycles are lost in the projection.
Another scenario which can result in incorrect interpretation is when exploring 
the long branches of the
star ensembles. The distance between two points adjacent in a long branch is not
necessarily the distance between them
in the high dimensional space. Rather, it represents the \textit{distance between the corresponding connected components}.
For example, say a point $p_1$ is connected to $p_2$ in the minimum spanning tree (and hence is an edge of the filtration). 
This does not guarantee that $p_1$ and $p_2$ will form an edge in the projection---the edge will be between the connected components 
corresponding to $p_1$ and $p_2$. Thus, the two points may be assigned to different branches of a star ensemble
depending on the strategy used during the projection.
Thus, this property must be considered when interpreting the layout.

\myparagraph{Layout interpretation guideline.}
Based on the above observations, we use the following guideline to
explore data using TopoMap for the remainder of this paper:
\squishlist
\item Use a density-based colormap to visualize the projection.
\item Start exploration by looking at centers of  
stars with high density. These typically represent clearly distinct clusters in 
the data.
\item Use low density stars to study ``uncommon behaviors".
\item Explore branches to analyze sparse clusters and outliers.
\squishend

\subsection{Case Study with Unlabeled Data}
\label{sec:case-studies}

There are several urban data sets available representing different facets of the city 
corresponding to its different properties. These are typically studied in isolation, 
and can sometimes result in missing out on interesting patterns resulting from
the interactions between these facets. For example, when analyzing just the count of
taxi trips, it is easy to observe that both Times Square as well as Penn Station
are identified as hot spots~\cite{doraiswamy@tvcg2014} almost throughout the day. 
Given that the former is a popular tourist attraction, while the latter is a 
transit hub, one would however expect differences in the way usage patterns
of these places change depending on other conditions. 

In this experiment, our goal is to see if such patterns do exist. To do so,
we generate high dimensional data sets by combining the NYC taxi data and the weather data as follows.
We divided two years (2014-2015) into hourly intervals. Then, given a location,
we consider the taxi pickups that happened within a 100~m radius of the location.
We then create one high dimensional point for every hourly interval 
having the following dimensions: count of taxi pickups, 
average fare, average distance, temperature, precipitation (rainfall), and wind speed. 
Thus, the data set corresponding to each location is a collection of 6D points.

We then projected the data corresponding to Times Square and Penn Station using TopoMap
and visually analyzed the patterns present in the projection
%
and analyzed the different stared ensembles by looking at 
the temporal distribution of the points forming these
clusters. 
The analysis of the data corresponding to Penn Station can be 
found in the supplemental material.

\myparagraph{Times Square.}
\autoref{fig:times-square} shows the results obtained for Times Square. 
In this scenario 
it is interesting to note that the most dense region (Figs.~\ref{fig:times-square}(b) and (c))
correspond to the summer months. This is in turn divided into two clusters: 
\autoref{fig:times-square}(b) 
corresponding to main part of the day (10~am to 6~pm), 
while \autoref{fig:times-square}(c) corresponds to night hours.
Similarly the winter months formed its own cluster~(\autoref{fig:times-square}(d)).
It was interesting to note that there was also a cluster with a smaller number of points
corresponding to the Spring and Autumn months (\autoref{fig:times-square}(e)).

A curious observation in the above cases was that none of these clusters included
the time period 4~am--8~am. We found these points for summer in a separate cluster
shown in \autoref{fig:times-square}(f). On further analysis, we found that
this is primarily because the taxi activity at these times was not only lower
than the other times, but that these trips also had longer distances and fares than normal.
Additionally, we also found that the points corresponding to periods when there 
was rainfall formed a less dense cluster among the outliers 
(\autoref{fig:times-square}(g)).


\begin{figure}[t]
	\centering
	\includegraphics[width=0.97\linewidth]{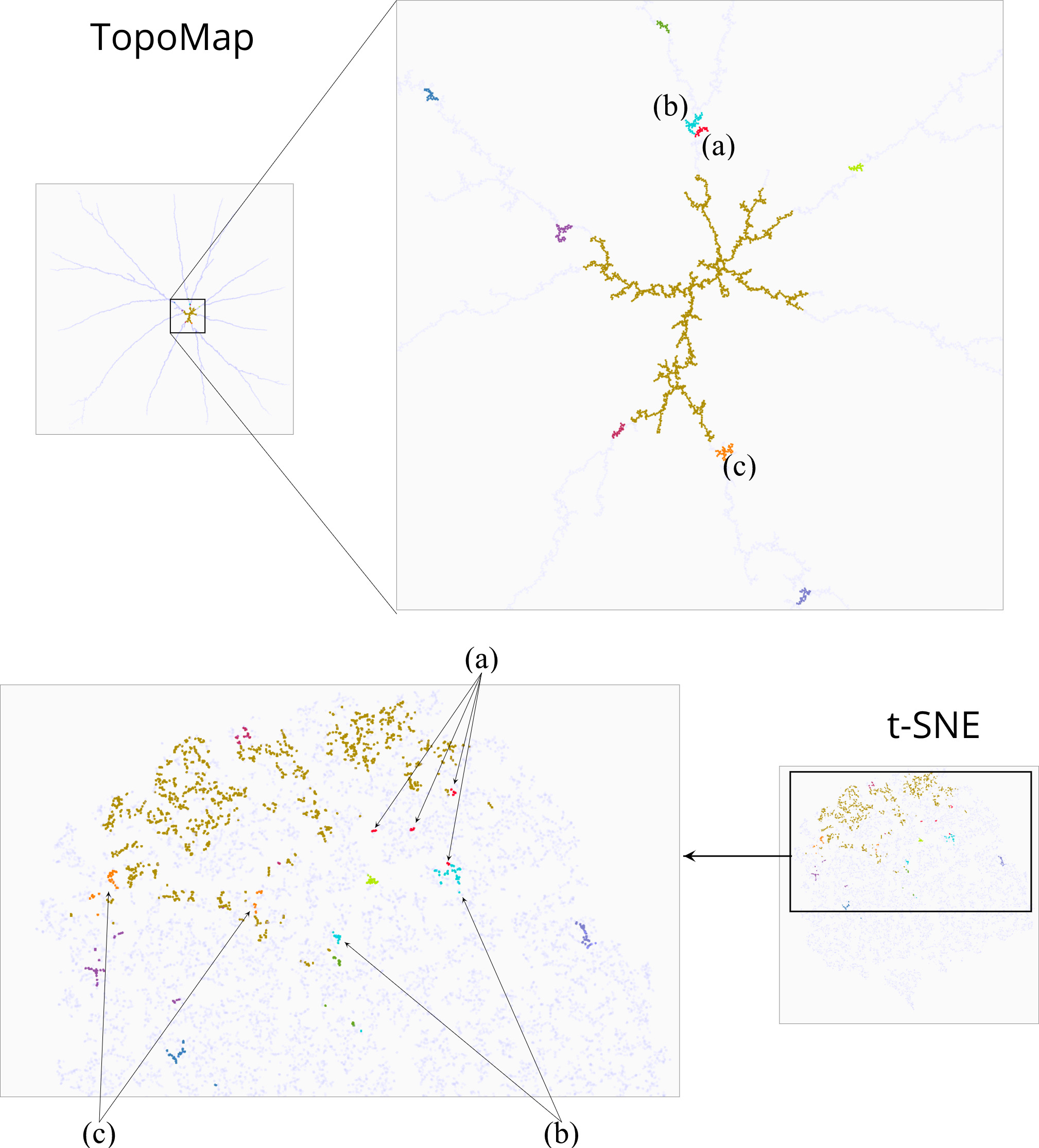}
	\vspace{-0.1in}
	\caption{
		Connected components (colored groups) guaranteed to exist in the high-dimensional space are broken apart by t-SNE.}
	\label{fig:probe}
	\vspace{-0.3in}
\end{figure}

\subsection{Probing Dissimilarity-based Projections}
\label{sec:probe}

Since TopoMap bears theoretical guarantees it can be used to probe other projection methods in order to further understand 
how those methods behave, specially regarding distortions and cluster preservation. To illustrate this,
consider the layouts in \autoref{fig:probe}. The layout on the top is the 
TopoMap 
projection of the urban data used in the previous section corresponding to Times Square. 
The highlighted points correspond to the 10 largest connected components 
obtained by stopping the topological filtration after adding 5000 topology 
changing edges. 
There is one large component (gold) and nine smaller ones 
highlighted in different colors.
Recall that if we apply this filtration in the high-dimensional space we would get exactly the same connected components.

The bottom image of \autoref{fig:probe} shows the result of projecting the same 
data using t-SNE.
The highlighted points here corresponds to the same components from the TopoMap layout. 
Notice how t-SNE spreads the large gold component around the layout. Even tightly connected components such as the 
ones indicated as (a), (b), and (c) in TopoMap layout are broken apart by t-SNE. This example reveals 
an important property of t-SNE, namely, clusters visualized in a t-SNE layout tend to correspond to pieces of clusters 
present in the high-dimensional data. With the help of TopoMap, one can realize where t-SNE is placing the different pieces of
a cluster.  Although experienced users are usually aware about this ``breaking cluster" property of t-SNE, we are not 
aware of any work capable of revealing the extent/intensity of this phenomenon.
Revealing this nature of t-SNE
is quite important, and can be considered as a side contribution of the present work helping to illustrate 
the potential of using TopoMap as an analytical tool.

\subsection{Discussions}
\label{sec:discussions}

\vspace{-0.1cm}
\myparagraph{Using alternate distance metrics.}
As shown in \autoref{sec:layoutinterp} 
(\autoref{fig:mnist}) TopMap can also be used with an alternate 
distance metric. This requires
computing the MST using this metric, in which case the running time 
for computing the MST degenerates to $O(N^2)$
due to the computation of the distance matrix.
Note that when another distance metric is used, the filtration in the projected
space is still preserved with respect to Euclidean distance in the visual space. 
This also makes it easier for the user to gauge the projection in the visualized space,
allowing for a comparison between the effect of using different distance metrics.

\myparagraph{Other 2D and 3D solutions.}
While our approach provides a solution
ensuring that the 0-dimensional homology is preserved, 
there can be other valid solutions as well.
Depending on the application, one can also trade-off 
preserving the persistence of outliers to preserving 
neighborhoods, or optimizing for a different property.
Similarly, it would be interesting to see how
the outliers would behave when moving to 3D.

\section{Conclusions}
In this paper we presented TopoMap, 
the first 
planar projection technique that is guaranteed to
preserve the homology of 0-cycles 
of the Rips filtration.
Evaluation of 
our approach using a variety of data sets demonstrated several key
properties that are desirable in a visual analytical tool: the layout is easy to understand
while its theoretical guarantees provide confidence to the users.
In the future, we would like to explore ways in which
1-cycles can be preserved as well in the projection. 
Analyzing the effectiveness of TopoMap to assist clustering mechanisms is another
direction we will pursue.


\myparagraph{Acknowledgments.}
This work was partially supported by the DARPA D3M program; Moore Sloan Data Science Environment at NYU; NSF awards CNS-1229185, CCF-1533564, CNS-1544753, CNS-1730396, CNS-1828576;
European Commission grant ERC-2019-COG \emph{``TORI''} (ref. 863464), CNPq-Brazil (303552/2017-4, 304301/2019-1); and  the S\~ao Paulo Research Foundation (FAPESP) - Brazil (2013/07375-0, 2016/04190-7, 2018/07551-6, 2018/24293-0). 
Any opinions, findings, and conclusions or recommendations expressed in this material are those of the authors and do not necessarily reflect the views of NSF and DARPA.


\bibliographystyle{abbrv}

\bibliography{paper}
\end{document}


\title{TopoMap: A 0-dimensional Homology Preserving Projection of High-Dimensional Data: Supplemental Material}

\author[1]{Harish Doraiswamy}
\author[2]{Julien Tierny}
\author[3]{Paulo J. S. Silva}
\author[4]{Luis Gustavo Nonato}
\author[1]{Claudio Silva}
\affil[1]{New York University}
\affil[2]{CNRS and Sorbonne Universit\'e}
\affil[3]{University of Campinas}
\affil[4]{University of Sao Paulo, Sao Carlos}
\date{}




\maketitle

\appendix
\renewcommand\thefigure{\thesection.\arabic{figure}} 

\section{Case Study with Unlabeled Data: Penn Station}

\begin{figure}[H]
	\centering
	\includegraphics[width=\linewidth]{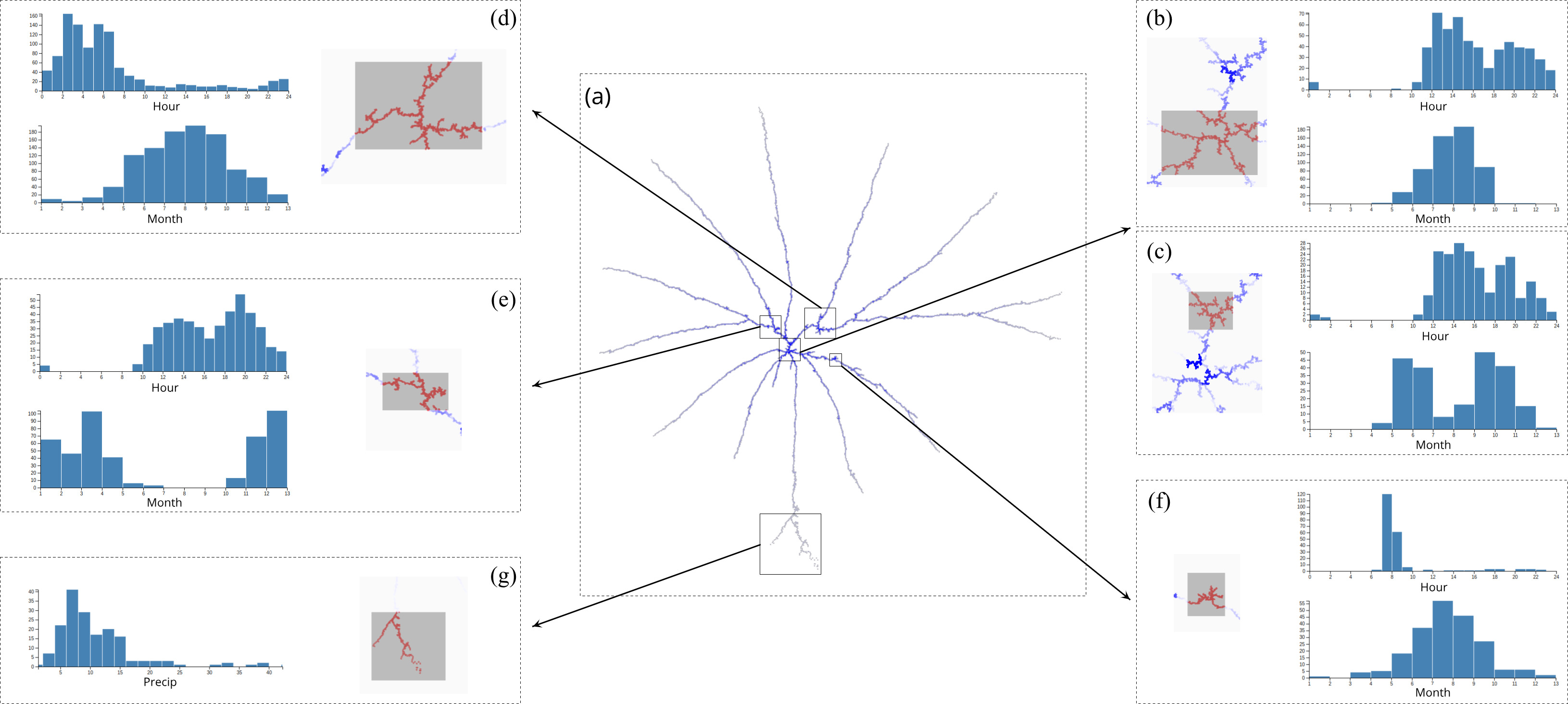}
	\caption{Analyzing Penn Station using TopoMap. 
		\textbf{(a)}~Projection obtained using TopoMap.
		\textbf{(b)--(g):}~Different clusters are selected and the temporal 
		distributions of the selected points visualized as a histogram. 
	}
	\label{fig:penn-station}
\end{figure}

%
Figure~\ref{fig:penn-station} shows the results obtained for Penn Station. 
Similar to Times Square, the dense clusters for Penn Station were also 
roughly divided based on the seasons and times of the day. However,
the actual division of times indicates a different kind of pattern
at this location. The clusters in Figure~\ref{fig:penn-station}(b)
corresponding to the time period 11~am to midnight during summer,
while Figure~\ref{fig:penn-station}(c) and (e) correspond to
the same time period during spring/autumn and winter respectively.
The cluster in Figure~\ref{fig:penn-station}(d) on the other hand 
corresponds to early hours of the day (midnight to 8~am).
While similar to Times Square, there was a separate cluster 
for the time period 7~am-9~am. Unlike the tourist location,
the cluster in this case was formed because the number of
taxi trips was significantly higher than in the other cases.
This is probably because of Penn Station being the transit hub
and the time period corresponds to peak hours. 
In fact, according to the US Census Bureau, over 1.6~million
workers commute daily to 
NYC\footnote{\url{
		https://www.census.gov/newsroom/press-releases/2013/cb13-r17.html}}.
Again, the periods of rainfall formed its own less dense cluster 
among the outliers (Figure~\ref{fig:penn-station}(g)).

\section{Probing UMAP using TopoMap}
%

\begin{figure}[t]
	\centering
	\includegraphics[width=0.6\linewidth]{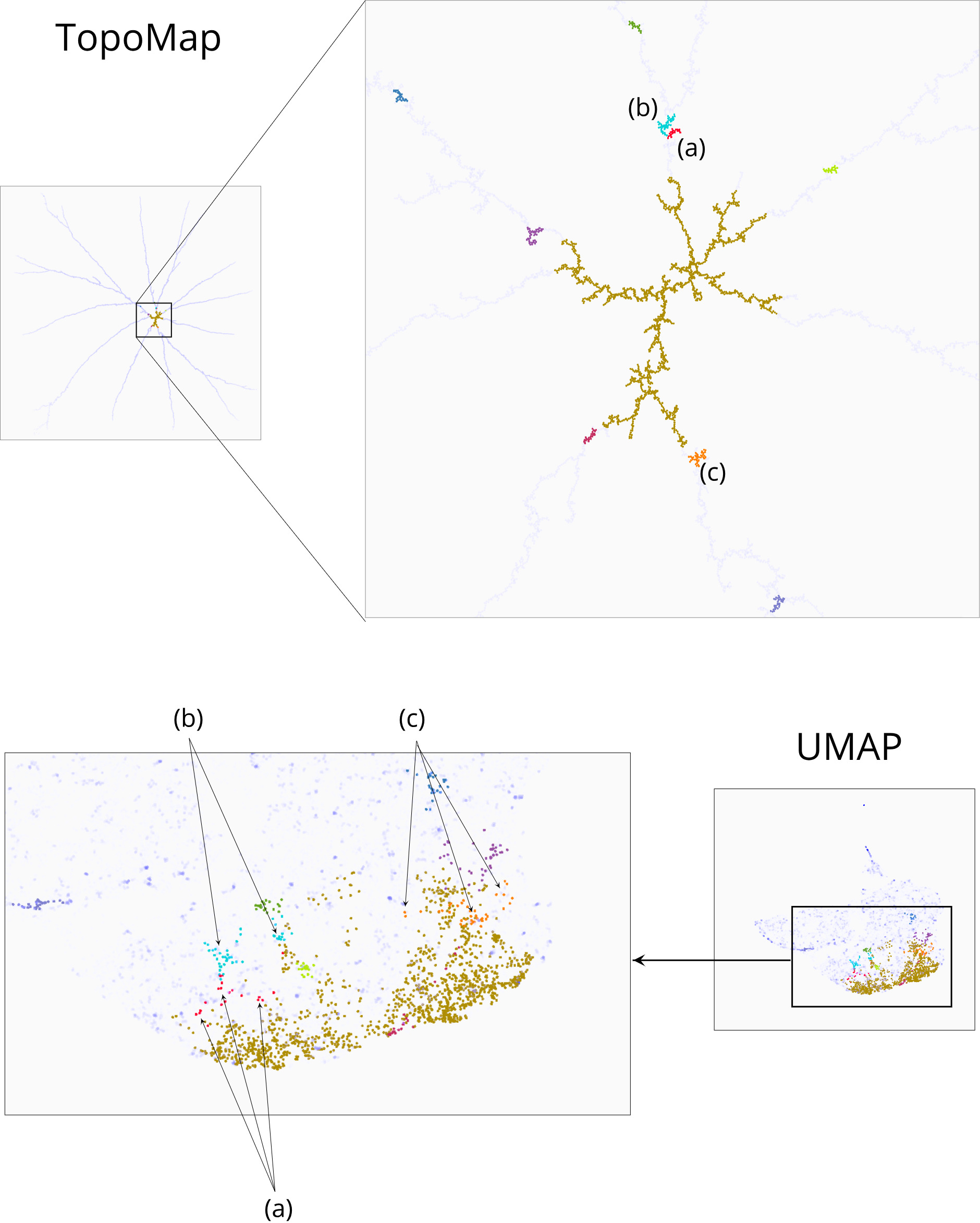}
	\caption{Probing UMAP with TopoMap. Similar to t-SNE (Figure~11), connected components (colored
		groups) are broken apart by UMAP as well.}
	\label{fig:umap-urban}
\end{figure}

%
Figure~\ref{fig:umap-urban} shows the projection obtained using UMAP on the data
used in Figure~11. As with t-SNE, not only 
the large gold component spreads around the layout, but even tightly connected
components such as the ones indicated as (a), (b), and (c) in TopoMap
layout are broken apart by UMAP.
From these examples, one can see that the visual results of UMAP 
are similar to that of t-SNE.

\bibliographystyle{abbrv-doi}
